\documentclass[journal]{IEEEtran}

\usepackage{cite}
\usepackage{url}
\usepackage{xcolor}
\usepackage{amsmath,amssymb,bm,amsfonts,amsthm,mathrsfs}
\usepackage{caption}
\usepackage{setspace}
\usepackage{bbm,multicol,multirow,makecell}
\usepackage{algorithm,algorithmic}
\usepackage{array,epsfig,graphicx}
\usepackage{stfloats}
\usepackage{lipsum}
\usepackage{setspace}
\usepackage{epstopdf}

\usepackage[T1]{fontenc}

\ifCLASSOPTIONcompsoc
  \usepackage[caption=false,font=normalsize,labelfont=sf,textfont=sf]{subfig}
\else
  \usepackage[caption=false,font=footnotesize]{subfig}
\fi

\ifCLASSOPTIONcaptionsoff
  \usepackage[nomarkers]{endfloat}
 \let\MYoriglatexcaption\caption
 \renewcommand{\caption}[2][\relax]{\MYoriglatexcaption[#2]{#2}}
\fi
\usepackage{url}

\theoremstyle{definition}
\newtheorem{proposition}{Proposition}
\newtheorem{theorem}{Theorem}
\newtheorem{lemma}{Lemma}

\newtheorem{assumption}{Assumption}

\allowdisplaybreaks[4]
\begin{document}

\title{Distributed Auto-Learning GNN for Multi-Cell Cluster-Free NOMA Communications}

\author{Xiaoxia~Xu, 
        Yuanwei~Liu,~\IEEEmembership{Senior~Member,~IEEE,}  
        Qimei~Chen,~\IEEEmembership{Member,~IEEE,}
        Xidong~Mu,~\IEEEmembership{Member,~IEEE,}
        and~Zhiguo~Ding,~\IEEEmembership{Fellow,~IEEE}

\thanks{
X. Xu and Q. Chen are with the School of Electronic Information, Wuhan University, Wuhan 430072, China (e-mail: \{xiaoxiaxu, chenqimei\}@whu.edu.cn).}
\thanks{Y. Liu and Xidong Mu are with the School of Electronic Engineering and Computer Science, Queen Mary University of London, London E1 4NS, U.K. (email: \{yuanwei.liu, xidong.mu\}@qmul.ac.uk).}
\thanks{Z. Ding is with the School of Electrical and Electronic Engineering, The University of Manchester, Manchester M13 9PL, U.K. (email: zhiguo.ding@manchester.ac.uk).}
}

\markboth{}%
{Shell \MakeLowercase{\textit{et al.}}: Bare Demo of IEEEtran.cls for IEEE Journals}

\maketitle

\begin{abstract}
    A multi-cell cluster-free NOMA framework is proposed, where both intra-cell and inter-cell interference are jointly mitigated via flexible cluster-free successive interference cancellation (SIC) and coordinated beamforming design.
    The joint design problem is formulated to maximize the system sum rate while satisfying the SIC decoding requirements and users' minimum data rate requirements.
    To address this highly complex and coupling non-convex mixed integer nonlinear programming (MINLP), a novel distributed auto-learning graph neural network (AutoGNN) architecture is proposed to alleviate the overwhelming information exchange burdens among base stations (BSs).
    The proposed AutoGNN can train the GNN model weights whilst automatically optimizing the GNN architecture, namely the GNN network depth and message embedding sizes, to achieve  communication-efficient distributed scheduling.
    Based on the proposed architecture, a bi-level AutoGNN learning algorithm is further developed to efficiently approximate the hypergradient in model training. 
    It is theoretically proved that the proposed bi-level AutoGNN learning algorithm can converge to a stationary point.
    Numerical results reveal that: 1) the proposed cluster-free NOMA framework outperforms the conventional cluster-based NOMA framework in the multi-cell scenario;
    and 2) the proposed AutoGNN architecture significantly reduces the computation and communication overheads compared to the conventional convex optimization-based methods 
    and the conventional GNNs with fixed architectures.

\end{abstract}


\IEEEpeerreviewmaketitle

\vspace{-0.2em}
\section{Introduction}
\IEEEPARstart{N}{ext}-generation wireless networks are envisioned to provide massive connectivity and high-quality transmissions for billions of bandwidth-hungry wireless devices in diversified scenarios \cite{IoE_2020, MA_2021}.
To meet these requirements, the concept of next-generation multiple access (NGMA) \cite{NGMA_2022} has been proposed to adaptively and intelligently provide wireless services for multiple users/devices given the limited radio resources.
Among others, the integration of multiple antenna technology with non-orthogonal multiple access (NOMA) is regarded as one of the most promising candidates for NGMA \cite{NGMA_2022}, 
which enables users to be served via the same orthogonal time/frequency/code resource while multiplexed in both the spatial and power domains.
However, conventional multi-antenna NOMA approaches have to group users into different clusters. 
By doing so, the intra-cluster and inter-cluster interference can be mitigated via the employment of successive interference cancellation (SIC) and the spatial beamforming \cite{MIMONOMA_2016,MIMONOMA_2017}.
Nevertheless, the effectiveness of traditional multi-antenna NOMA approaches rely on specific scenarios, which may not always hold due to the channel randomness.
To address this issue, a generalized multi-antenna NOMA transmission framework was proposed in \cite{ClusterFree_2022} with a novel concept of cluster-free SIC. By breaking the limitation of sequentially carrying out SIC within each cluster, 
the proposed cluster-free NOMA is capable to achieve efficient interference suppression and high communication performance in various scenarios.

Recall the fact that network densification is a key enabler for enhancing the network capacity and providing ubiquitous access. 
On the road to NGMA, one of the most fundamental and practical problem is how to design efficient multi-cell NOMA communications. 
Despite providing an enhanced transmission flexibility, the investigations on cluster-free NOMA communication are still in an early stage.  
In particular, its effectiveness and application in more practical multi-cell networks is still not investigated. 
To fill this gap, we propose the multi-cell cluster-free NOMA framework in this paper, where both intra-cell and inter-cell interference should be mitigated compared to \cite{ClusterFree_2022}. 
In light of this, the coordinated scheduling of base stations (BSs) is crucial.
However, this is usually highly computational complexity, and necessitates sharing locally available channel state information (CSI) among BSs.
To reduce the computational complexity and relieve overwhelming information exchange overheads, the centralized scheduling method for the single-cell network is no longer applicable, which makes it urgent to design efficient distributed scheduling methods.

\subsection{Prior Works}
Existing multi-cell distributed scheduling methods can be loosely divided into two categories, i.e., the conventional optimization-based and the emerging learning-based methods.
\subsubsection{Conventional optimization-based distributed scheduling}
The authors of \cite{MIMONOMA_CB_2017} developed two interference alignment based coordinated beamforming schemes for two-cell multiple-input multiple-output (MIMO)-NOMA networks, which successfully deal with inter-cell interference and increase the throughput of cell-edge users.
The authors of \cite{CoMPNOMA_power_2018} investigated a Karush-Kuhn-Tucker based distributed optimization method in coordinated beamforming (CoMP)-NOMA networks, where BSs locally optimize power allocation strategies.
The authors of \cite{MISONOMA_power_2020} investigated distributed joint user grouping, beamforming, and power control strategies to minimize the power consumption of multi-cell multiple-input single-output (MISO)-NOMA networks through zero-forcing beamforming, semiorthogonal user selection, and power consumption oriented user grouping.
Additionally, to maximize the energy efficiency under imperfect SIC, the authors of \cite{MIMONOMA_THzEE_2020} developed a distributed alternating direction method of multipliers (ADMM) for coordinated power allocation in a downlink heterogeneous Terahertz MIMO-NOMA network.
Moreover, by considering both perfect and imperfect CSI, the authors of \cite{MIMONOMA_WPTEE_2021} developed a distributed ADMM-based resource allocation algorithm to maximize the energy efficiency for a massive MIMO-NOMA network.

\subsubsection{Emerging learning-based distributed scheduling}
Deep learning (DL) has been widely considered as a promising paradigm for distributed scheduling \cite{MPGNN_2021,UnfoldingGNN_2021,GEMAL_2022,HetGNN_2022}. 
However, conventional deep neural networks (DNNs), e.g., multi-layer perceptron (MLP) and convolutional neural network (CNN), are originally inherited from computation vision domain, which usually suffer poor generalizations and scalability when being employed in wireless systems \cite{MPGNN_2021}. 
While specific solutions such as Learning to Optimize for Resource Management (LORM) \cite{LORM} have been proposed to handle mixed-integer nonlinear programming (MINLP) in wireless systems, they are typically centralized scheduling methods.
As a remedy, graph neural network (GNN) provides a structural learning framework for distributed interference (scheduling), which can capture the underlying dependencies between wireless nodes. 
By embedding structural features into messages and passing the learned messages between distributed wireless nodes, distributed scheduling can be locally predicted by each wireless node to achieve efficient coordination \cite{GNN_2021,MPGNN_2021}.
In \cite{MPGNN_2021}, the authors identified the effectiveness of message passing GNN for solving the distributed power control and beamforming problems, and theoretically analyzed its permutation-equivalence property, scalability, and generalization ability.
Alternatively, the authors of \cite{UnfoldingGNN_2021} unfolded a power allocation enabled iterative weighted minimum mean squared error (WMMSE) algorithm with a distributed GNN architecture, which achieves higher robustness and generalizability in unseen scenarios.
In reconfigurable intelligent surface (RIS) aided terahertz massive MIMO-NOMA networks, the authors of \cite{GEMAL_2022} integrated the graph neural network into distributed multi-agent deep reinforcement learning architecture to facilitate information interaction and coordination.
Moreover, the authors of \cite{HetGNN_2022} learned a distributed heterogeneous GNN over wireless interference graph with a parameter sharing scheme, which enables more efficient scheduling than homogeneous GNNs.

\subsection{Motivations}
While the aforementioned investigations have been devoted to conventional multi-cell NOMA, there is still a paucity of research on multi-cell cluster-free NOMA.
Furthermore, the above optimization-based  and learning-based distributed scheduling methods have their own demerits to deal with the distributed optimization in multi-cell networks.
\begin{itemize}
  \item Conventional optimization-based distributed scheduling methods typically require a large number of iterations to converge \cite{DecADMM_2011}, 
      which is inapplicable for tasks with low-latency requirements.
      Moreover, when dealing with mixed-integer optimization variables, the obtained results are highly sensitive to initialized parameters, which need to be carefully and manually configured for different scenarios.
      In a nutshell, the slow convergence and the hand-engineered initialized parameters require frequent information exchange among BSs, leading to overwhelming communication and computation burdens in practice.
     
  \item Emerging learning-based distributed scheduling methods can directly learn the mapping from agents' observations to the optimal solutions, 
  thus overcoming the initialized-parameter dependence. 
  Furthermore,  
  they can achieve real-time distributed scheduling by performing low-complexity forward propagation over limited neural layers that are trained to approximate desirable optimization solutions. 
  However, conventional DNNs are awkward to exploit structural features and lack generalization capabilities. 
  Although GNNs can compensate for several weaknesses of DNNs, they still suffer from predefined hyperparameters and inflexible fixed architectures, which can result in inefficiency for distributed scheduling.
\end{itemize}

Against the above background, this paper investigates the sum rate maximization for multi-cell NOMA networks by jointly optimizing the multi-cell coordinated beamforming and cluster-free SIC operations. 
The joint optimization problem is a highly coupling and complex non-convex MINLP.
To overcome the shortcomings of both conventional optimization-based and learning-based distributed scheduling methods, 
we propose a novel distributed automated-learning graph neural network (AutoGNN) framework, where distributed BSs can cooperatively infer their local optimization variables via a learning-based lightweight mechanism.
Inspired by automated machine learning (AutoML) \cite{AutoML_2019}, auto-learning GNNs have been investigated in machine learning domain recently, 
which focus on minimizing specific loss functions without considering computation and communication burdens for distributed inference \cite{AutoGEL}. 
However, since relieving these burdens is a key incentive to employ GNNs in wireless systems, it is crucial to propose cost-efficient auto-learning GNNs for wireless systems. 
Motivated by this, this paper proposes AutoGNN, a self-optimized GNN architecture that enables a novel communication-efficient distributed scheduling paradigm. 
When training the GNN model, AutoGNN adaptively learns desirable GNN architecture parameters, namely GNN network depth and message embedding sizes,  
which can flexibly reduce computational and information exchange overheads whilst optimizing the system performance. 

\vspace{-0.2cm}
\subsection{Contributions}
The main contributions of this work can be summarized as follows.
\begin{itemize}
  \item We propose a multi-cell cluster-free NOMA framework, 
  where both intra-cell and inter-cell interference can be efficiently mitigated via joint SIC operations and coordinated beamforming. 
  The objective function is formulated to maximize the system sum rate while satisfying the SIC decoding requirements, which is a highly coupling and complex MINLP problem.
  \item We propose a novel communication-efficient learning architecture, namely AutoGNN, 
  for learning distributed multi-cell coordinated beamforming and cluster-free SIC optimization.
  Different from conventional GNNs with fixed architectures, the proposed AutoGNN architecture automatically learns the desirable GNN network depths and message embedding sizes.
   Hence, the communication and computation overheads can be adaptively reduced.
  \item We develop a bi-level AutoGNN learning algorithm to jointly train the GNN model weights and architecture parameters, which can approximate the hypergradient efficiently. 
  Moreover, we analyze the upper bound of the approximation error and theoretically demonstrate that the bi-level AutoGNN learning algorithm  can converge to the stationary point.
  \item  Numerical results verify that the proposed multi-cell cluster-free NOMA framework outperforms cluster-based NOMA framework in various scenarios. 
  Moreover, the proposed AutoGNN can effectively and adaptively reduce communication overheads compared to conventional fixed-architecture GNNs and distributed ADMM algorithm.
\end{itemize}

The rest of this paper is organized as follows. 
Section II describes the downlink multi-cell cluster-free multi-antenna NOMA framework and formulates the sum rate maximization problem. 
A communication-efficient AutoGNN architecture is proposed in Section III. 
Next, a bi-level AutoGNN learning algorithm is developed in Section IV. 
In Section V, numerical results are presented to verify the effectiveness of the proposed framework and learning algorithms. Finally, Section VI concludes the paper.

\section{System Model and Problem Formulation}
In this section, we first introduce a downlink multi-cell cluster-free NOMA framework, 
and then discuss the problem formulation to efficiently suppress both intra-cell and inter-cell interference.

\subsection{Downlink Multi-Cell Cluster-Free NOMA Framework}

\begin{figure*}[!htbp]
  \vspace{-1em}
  \centering
  \includegraphics[width=1\textwidth]{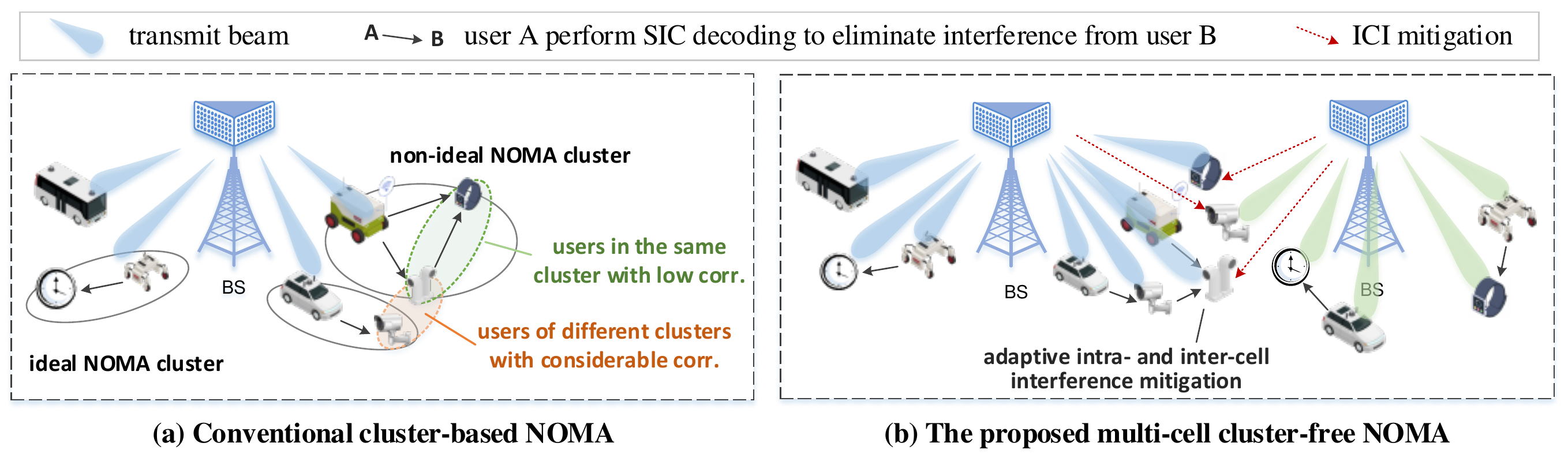}\\
  \caption{Illustration of the proposed downlink multi-cell cluster-free NOMA framework. }\label{fin_model}
  \vspace{-0.2em}
\end{figure*}

As illustrated in Fig. \ref{fin_model}, we propose a downlink coordinated multi-cell multi-antenna cluster-free NOMA framework, 
which consists of a set $\mathcal{M}$ of $M$ densely deployed BSs connected via wireless backhaul.
Each BS equips $N_{\mathrm{T}}$ antennas to serve a set $\mathcal{K}_m$ of $K_m$ single-antenna users within its coverage.
Without loss of generality, we assume $K = K_1 = K_2 = ... = K_M$.
Note that the proposed framework can be employed in both underloaded ($K\leq N_{\mathrm{T}}$) system and overloaded ($K >  N_{\mathrm{T}}$) system.
Here, users served by each BS $m$ are ordered according to the ascending order of their data channel gains.
Define the transmit beamforming matrix of BS $m$ as $\mathbf{W}^{m}=\left[\mathbf{w}_{1}^{m},\mathbf{w}_{2}^{m},...,\mathbf{w}_{K}^{m}\right]\in\mathbb{C}^{N_{\mathrm{T}}\times K}$, where $\mathbf{w}_{k}^{m}\in\mathbb{C}^{N_{\mathrm{T}}\times 1}$ denotes the dedicated beamforming vector from BS $m$ to user $k$.

The received signal at user $k \in \mathcal{K}_{m}$ can be represented by
\vspace{-0.2em}
\begin{equation}\label{Signal}
    \begin{split}
    y_{k}^{m}\!&=\!\underset{\text{desired signal}}{\underbrace{\left|\mathbf{h}_{mk}^{m}\mathbf{w}_{k}^{m}\right|^{2}\!\sqrt{s_{k}^{m}}}}
    +\underset{\text{intra-cell interference}}{\underbrace{\sum_{u\ne k}\!\left|\mathbf{h}_{mk}^{m}\mathbf{w}_{u}^{m}\right|^{2}\!\!\sqrt{s_{u}^{m}}}}
    \\&+\underset{\text{inter-cell interference}}{\underbrace{\sum_{n\ne m}\sum_{u\in\mathcal{K}_{n}}\!\!\left|\mathbf{h}_{mk}^{n}\mathbf{w}_{u}^{n}\right|^{2}\!\sqrt{s_{u}^{n}}}}+\underset{\text{noise}}{\underbrace{z_{k}^{m}}},
    ~\forall k\in\mathcal{K}_{m},~m\in\mathcal{M},
    \end{split}
\end{equation}
\vspace{-0.2em}

\noindent where $\mathbf{h}_{mk}^{n}\in\mathbb{C}^{1\times N_{\mathrm{T}}}$ denotes the channel from BS $n$ to the $k$-th user served by BS $m$, and $\mathbf{h}_{mk}^{m}\in\mathbb{C}^{1\times N_{\mathrm{T}}}$ signifies the channel from BS $m$ to the $k$-th user served by BS $m$.

To reduce both the inter-cell and intra-cell interference, we jointly employ the multi-cell coordinated beamforming and the SIC to transmit and decode users' signals as shown in Fig. \ref{fin_model}. Specifically,
we introduce a cluster-free SIC scheme \cite{ClusterFree_2022}, where users can flexibly mitigate intra-cell interference unimpeded by predefined user clusters.

We specify the cluster-free SIC operations for each user $k\in\mathcal{K}_{m}$ with a binary vector $\bm{\beta}_{k}^{m} = \left[\beta_{1k}^{m},\beta_{2k}^{m},...\beta_{Kk}^{m}\right]^T$, where $\beta_{ik}^{m}\in\{0,1\}$ indicates whether user $i\in\mathcal{K}_{m}$ would carry out SIC to decode the signal of user $k\in\mathcal{K}_{m}$, $i\ne k$, before decoding its own signal.
The achievable rate of SIC decoding and downlink transmission can be modelled as follows.

\subsubsection{SIC decoding rate}
The interference $\mathrm{Intf}_{ik}^{m}\left(\bm{\beta}^{m},\mathbf{W}\right)$ for user $i$ to decode the signal of user $k$, $i\ne k$, can be formulated by \cite{ClusterFree_2022}
\vspace{-0.1cm}
\begin{equation}\label{Intf_decoding1}
\begin{split}
& \mathrm{Intf}_{ik}^{m}\left(\bm{\beta}^{m},\mathbf{W}\right) =
\underset{\text{intra-cell interference from weaker users}}{\underbrace{\sum\limits _{u<k}\left(1-\beta_{iu}^{m}+\beta_{iu}^{m}\beta_{uk}^{m}\right)\left|\mathbf{h}_{mi}^{m}\mathbf{w}_{u}^{m}\right|^{2}}} \\&
+ \underset{\text{intra-cell interference from stronger users}}{\underbrace{\sum\limits _{u>k}\left(1-\beta_{iu}^{m}\beta_{ku}^{m}\right)\left|\mathbf{h}_{mi}^{m}\mathbf{w}_{u}^{m}\right|^{2}}} 
+ \mathrm{ICI}_{i}^{m}\left(\mathbf{W}\right),
\\& \hspace{2em} ~\forall i\ne k, ~ i,k\in\mathcal{K}_{m}, ~ m\in\mathcal{M},
\end{split}
\vspace{-0.2cm}
\end{equation}
where $\bm{\beta}^{m}=\left[\bm{\beta}_{1}^{m},\bm{\beta}_{2}^{m}...,\bm{\beta}_{K}^{m}\right]$ and $\mathbf{W}=\left[\mathbf{W}^{1},\mathbf{W}^{2},...,\mathbf{W}^{M}\right]$ denote the stacked matrices, 
and $\mathrm{ICI}_{i}^{m}\left(\bm{\beta}^{m},\mathbf{W}\right)=\sum\limits _{n\ne m}\sum_{u\in\mathcal{K}_{n}}\left|\mathbf{h}_{mi}^{n}\mathbf{w}_{u}^{n}\right|^{2}$ denotes the inter-cell interference (ICI) suffered by user $i$.
Hence, when user $i$ decoding user $k$'s signal, the received SINR $\gamma_{ik}^{m}$ can be expressed as
\vspace{-0.2cm}
\begin{equation}\label{SINR_decoding1}
\gamma_{ik}^{m} =
\frac{\left|\mathbf{h}_{mi}^{m}\mathbf{w}_{k}^{m}\right|^2}{\mathrm{Intf}_{ik}^{m}\left(\bm{\beta}^{m},\mathbf{W}\right)+\sigma^2},
~\forall i\ne k, ~ i,k\in\mathcal{K}_{m}, ~ m\in\mathcal{M}.
\vspace{-0.2cm}
\end{equation}
Therefore, the corresponding SIC decoding rate can be derived by $r_{ik}^{m} \!=\! \log_{2}\left(1+\gamma_{ik}^{m}\right)$.

\subsubsection{Transmission rate}
When user $k$, $\forall k\in\mathcal{K}_{m}$, decoding its own signal, the interference can be expressed as
\vspace{-0.1cm}
\begin{equation}\label{Intf_decoding2}
\mathrm{Intf}_{kk}^{m}\left(\bm{\beta}^{m},\mathbf{W}\right)
=\underset{\text{intra-cell interference after SIC}}{\underbrace{\sum\limits _{u\ne k}\left(1-\beta_{ku}^{m}\right)\left|\mathbf{h}_{mk}^{m}\mathbf{w}_{u}^{m}\right|^{2}}}
+\underset{\text{inter-cell interference}}{\underbrace{\mathrm{ICI}_{k}^{m}\left(\mathbf{W}\right)}}.
\end{equation}
The corresponding transmission rate of user $k$ can be computed by $r_{kk}^{m} \!=\! \log_{2}\!\left(1\!+\!\gamma_{kk}^{m}\right) \!=
\! \log_{2}\!\left(1+\frac{\left|\mathbf{h}_{mk}^{m}\mathbf{w}_{k}^{m}\right|^{2}}{\mathrm{Intf}_{kk}^{m}\left(\bm{\beta}^{m},\mathbf{W}\right)+\sigma^2}\right)$.

To correctly decode the intended signal of user $k$, the received SINR for user $i$ to decode user $k$'s signal should be larger than or equal to the received SINR of user $k$ to decode its own signal, $\forall \beta_{ik}^{m}=1$ \cite{SIC_2016}.
Owing to the SIC decoding constraint, the effective data rate $R_{k}^{m}$ for user $k$ is given by
\vspace{-0.2em}
\begin{equation}\label{SICConstraint}
R_{k}^{m}=\min_{i\in\mathcal{K}_{m}} \left\{ \frac{1}{\beta_{ik}}r_{ik}^{m} \right\}, ~ \forall  k\in\mathcal{K}_{m}, m \in\mathcal{M}.
\vspace{-0.2em}
\end{equation}
By considering the physical meaning, the above rate expression can be further rewritten as the following continuous form that is more tractable for learning algorithms: 
\vspace{-0.2em}
\begin{equation}
R_{k}^{m} \!= \!
 \min_{i\in\mathcal{K}^{m}}\! \bigg\{\! \beta_{ik}^{m} r_{ik}^{m} \!+\! \left(1\!-\!\beta_{ik}^{m}\right)r_{kk}^{m}\!\bigg\}, 
      ~ \forall k \in \mathcal{K}_{m}, ~ m\in\mathcal{M}.
\vspace{-0.2em}
\end{equation}

\vspace{-0.5em}
\subsection{Problem Formulation}
Based on the proposed multi-cell cluster-free NOMA framework, we aim to maximize the system sum rate through jointly optimizing coordinated beamforming $\mathbf{W}$ and SIC operations $\bm{\beta}$, 
subject to the SIC decoding constraints and users' minimal data rate requirements, which can be formulated as
\vspace{-0.2em}
\begin{subequations}\label{P0}
\begin{align*}
\mathcal{P}_{0}:~ & \max_{\bm{\beta},\mathbf{W}} ~
\sum\limits_{m\in\mathcal{M}}\sum\limits_{k\in\mathcal{K}_{m}} R_{k}^{m}  \label{P0_obj} \tag{\ref{P0}{a}}
\\ {\text{s.t.}}~~~ &
R_{k}^{m}\ge R_{k}^{m,\mathrm{\min}}, ~\forall k \in \mathcal{K}_{m},  m \in\mathcal{M}, \label{constraint_rate} \tag{\ref{P0}{b}}
\\&
\sum\limits_{k\in\mathcal{K}_m} \left\|\mathbf{w}_{k}^{m}\right\|^2 \le P^{\max}, ~ \forall m \in\mathcal{M},\label{constraint_power} \tag{\ref{P0}c}
\\&
\beta_{ik}^{m} + \beta_{ki}^{m} \le 1, ~\forall i\ne k,  i,k \in \mathcal{K}_{m}, m \in\mathcal{M}, \label{constraint_Decoding} \tag{\ref{P0}d}
\\&
\beta_{ik}^{m} \in\{0, 1\}, ~ \forall i,k \in \mathcal{K}_{m}, m \in\mathcal{M}, \label{constraint_bin} \tag{\ref{P0}e}
\end{align*}
\vspace{-0.5cm}
\end{subequations}

\noindent where constraint \eqref{constraint_rate} guarantees the minimum data rate requirement $R_{k}^{m,\min}$ of each user $k\in\mathcal{K}_{m}$,
and \eqref{constraint_power} ensures that the transmission power of each BS does not exceed $P^{\max}$. 
Constraint \eqref{constraint_Decoding} indicates that user $i$ and user $k$, $i\ne k$, cannot mutually carry out SIC.
Intuitively, $\mathcal{P}_{0}$ is a highly coupling and complex non-convex MINLP, which is an NP-hard problem that is challenging to be optimally solved in a centralized way.
To reduce computation complexity as well as relieve information exchange burdens, 
efficient distributed scheduling methods should be investigated to obtain desirable solutions.

\section{Communication-Efficient AutoGNN Architecture}
In this section, we propose a novel AutoGNN architecture for achieve communication-efficient distributed scheduling in multi-cell cluster-free NOMA networks.
We first model the proposed multi-cell cluster-free NOMA framework as a distributed communication graph.
Thereafter, a novel communication-efficient AutoGNN architecture is proposed to overcome the inefficiency of conventional message passing GNN built on the fixed architecture. 

\subsection{Distributed Communication Graph Model}
The proposed multi-cell cluster-free NOMA framework can be modelled as a distributed communication graph, defined as $\mathcal{G}=\left\{ \mathcal{M},\mathcal{E},\mathbf{O},\mathbf{X}\right\}$.
Specifically, $\mathcal{G}$ is a directed graph, where all BSs are modelled as a set of nodes $\mathcal{M}$, and the interplay effects among BSs are represented by a set of edges $\mathcal{E}$.
Define $E_{mn}\in\mathcal{E}$ as the edge from node $m$ to node $n$, which is an outbound edge of node $m$ and an inbound edge of node $n$.
Let $\mathcal{N}_{m}^{\mathrm{out}}$ ($\mathcal{N}_{m}^{\mathrm{in}}$) denote the set of nodes connecting with node $m$ through an outbound (inbound) edge of node $m$, and $N_{m}^{\mathrm{out}}$ and $N_{m}^{\mathrm{in}}$ are the cardinality of sets $\mathcal{N}_{m}^{\mathrm{out}}$ and $\mathcal{N}_{m}^{\mathrm{in}}$, respectively.
Moreover, $\mathbf{O}=\left[\mathbf{O}_{1},\mathbf{O}_{2},...,\mathbf{O}_{M}\right]$ and $\mathbf{X}=\left[\mathbf{X}_{1},\mathbf{X}_{2},...,\mathbf{X}_{M}\right]$ denote the joint observations and hidden states, where $\mathbf{O}_{m}$ and $\mathbf{X}_{m}$ denote the local observation and local hidden states at BS $m$, respectively.
The local observation $\mathbf{O}_{m}$ is partially observable by BS $m$, which consists of the node feature $\mathbf{O}_{m}^{\mathrm{N}}$ and the edge feature $\left\{\mathbf{O}_{mn}^{\mathrm{E}}\right\}_{n \in \mathcal{N}_{m}^{\mathrm{out}}}$.

Based on the directed graph model, we can straightforwardly model the data channels from each BS $m$ to its serving users as its node feature, and depict the interference channels from BS $m$ to the device served by other BSs as its edge features, which can be expressed as
\vspace{-0.1em}
\begin{equation}\label{node_feature}
\mathbf{O}_{m}^{\mathrm{N}}=\mathbf{H}_{mm}=\left[\left\{ \mathbf{h}_{mk}^{m}\right\}_{k\in\mathcal{K}^{m}}\right], ~ \forall m \in \mathcal{M},
\vspace{-0.1em}
\end{equation}
\begin{equation}\label{edge_feature}
\mathbf{O}_{mn}^{\mathrm{E}}=\mathbf{H}_{mn}=\left[\left\{ \mathbf{h}_{nk}^{m}\right\}_{k\in\mathcal{K}^{n}}\right], ~ \forall m\ne n, n \in \mathcal{N}_{m}^{\mathrm{out}},
\vspace{-0.1em}
\end{equation}
where $\mathbf{H}_{mm}\in\mathbb{C}^{N_{\mathrm{T}}\times K}$ collects the data channels from BS $m$ to its serving users, and $\mathbf{H}_{mn}\in\mathbb{C}^{N_{\mathrm{T}}\times K}$ stacks the interference channels from BS $m$ to users served by the neighbouring BS $n\in \mathcal{N}_{m}^{\mathrm{out}}$, respectively.
Moreover, the local hidden states at each BS $m$ can be initialized by the input node feature, i.e.,
$\mathbf{X}_{m}^{0}=\mathbf{O}_{m}^{\mathrm{N}}$.
Based on the distributed communication graph model, we introduce the conventional message passing GNN and the proposed AutoGNN architecture as follows.

\subsection{Conventional Message Passing Graph Neural Network} \label{Sec_MPGNN}
Conventional message passing GNN \cite{MPGNN_2021} can eliminate the parameter initialization dependence
and avoid the slow convergence of iterative optimization algorithms.
Following the principle of DNNs, GNN exploits a multi-layer structure.
Define $L$ as the number of GNN layers, and $\bm{\theta}=\left[\bm{\theta}^{(1)},\bm{\theta}^{(2)},...,\bm{\theta}^{(L)}\right]$ as the set of weights of the entire GNN model.
Each GNN layer $l$ includes a distributed message passing process to achieve agents' coordination, which consists of three steps, namely the message embedding, the message aggregation, and the message combination.
The detailed process can be illustrated as follows.

\textbf{(i) Message embedding.}
At each layer $l$, agent $m$ embeds the local hidden state $\mathbf{X}_{m}^{(l)}$ and the outbound edge feature $\mathbf{O}_{mn}^{\mathrm{E}}$ to obtain a message vector $\mathbf{u}_{mn}^{(l)}=\left[u_{mn1}^{(l)},u_{mn2}^{(l)},...,u_{mnD^{\mathrm{E}}}^{(l)}\right]\in\mathbb{R}^{D^{\mathrm{E}}\times1}$, where $D^{\mathrm{E}}$ represents the embedding size of $\mathbf{u}_{mn}^{(l)}$.
Thereafter, agent $m$ would send the outbound embedded message $\mathbf{u}_{mn}^{(l)}$ to agent $n\in\mathcal{N}_{m}^{\mathrm{out}}$, and then receives the inbound embedded message $\mathbf{u}_{nm}^{(l)}$ from agent $n\in\mathcal{N}_{m}^{\mathrm{in}}$.
The embedded message $\mathbf{u}_{mn}^{(l)}$ from agent $m$ to agent $n$ can be obtained by the local encoder $\phi_{\mathrm{E}}^{l}$ as
\vspace{-0.2em}
\begin{equation}\label{FGNN_Embed}
\mathbf{u}_{mn}^{(l)}= \phi_{\mathrm{E}}^{l}\left(\mathbf{X}_{m}^{(l-1)},\mathbf{O}_{mn}^{\mathrm{E}};\bm{\theta}_{\mathrm{E}}^{(l)}\right), ~\forall n \in \mathcal{N}_{m}^{\mathrm{out}},
\vspace{-0.2em}
\end{equation}
where $\phi_{\mathrm{E}}^{l}\left(\cdot;\bm{\theta}_{\mathrm{E}}^{(l)}\right)$ denotes the local embedding function at layer $l$, which is implemented as MLPs parameterized by $\bm{\theta}_{\mathrm{E}}^{(l)}$.

\textbf{(ii) Message aggregation.}
After receiving the embedded messages $\mathbf{u}_{nm}^{(l)}$ from neighbouring agents $n\in\mathcal{N}_{m}^{\mathrm{in}}$, agent $m$ aggregates the received messages $\mathbf{u}_{nm}^{(l)}, \forall n\in\mathcal{N}_{m}^{\mathrm{in}},$ as
\vspace{-0.2em}
\begin{equation}\label{FGNN_Agg}
\overline{\mathbf{u}}_{m}^{(l)} = \phi_{\mathrm{A}}\left(\left\{ \mathbf{u}_{nm}^{(l)}\right\} _{\forall n\in\mathcal{N}_{m}^{\mathrm{in}}};\bm{\theta}_{\mathrm{A}}^{(l)}\right),
\vspace{-0.2em}
\end{equation}
where $\phi_{\mathrm{A}}$ denotes a permutation-invariant function, such as $\mathrm{mean}(\cdot)$, $\mathrm{sum}(\cdot)$, and $\mathrm{max}(\cdot)$.

\textbf{(iii) Message combination.}
Given the combination function $\phi_{\mathrm{C}}$, the local hidden state at agent $m$ can be updated by
\vspace{-0.2em}
\begin{equation}\label{FGNN_Comb}
\mathbf{X}_{m}^{(l)}=\psi_{\mathrm{C}}\left(\mathbf{X}_{m}^{(l-1)},\overline{\mathbf{u}}_{m}^{(l)};\bm{\theta}_{\mathrm{C}}^{(l)}\right).
\vspace{-0.2em}
\end{equation}

Define the optimization variables  predicted by GNN as $\mathbf{Z}=\left[\mathbf{Z}_{1},\mathbf{Z}_{2},...,\mathbf{Z}_{M}\right]$, with $\mathbf{Z}_{m}=\left\{\bm{\beta}^{m},\widetilde{\bm{\beta}}^{m},\mathbf{W}^{m}\right\}$ being the local optimization variables at BS $m$.
The optimization variables can be obtained via a fully connected (FC) layer, which can be expressed as
\vspace{-0.2em}
\begin{equation}\label{FGNN_Output}
\mathbf{Z}_{m}
= \psi_{\mathrm{FC}}\left(\mathbf{O}_{m}^{\mathrm{N}}, \mathbf{X}_{m}^{(L)}\right),
\vspace{-0.2em}
\end{equation}
where $\varphi_{\mathrm{FC}}(\cdot)$ denotes the FC layer function.

\subsection{Communication-Efficient AutoGNN Architecture}
To accommodate various communication scenarios, the static GNN architectures should be artificially and empirically designed for different environments. However, it is generally time-consuming, laborious, and error-prone to search for the optimal neural network architecture and hyperparameters.
Hence, neural architecture searching (NAS) \cite{NAS_2019,DARTS, iDARTS, AutoML_2019} has been proposed as a promising AutoML paradigm to automate neural network designs.
Inspired by NAS, we propose a novel communication-efficient AutoGNN architecture, which automates the architecture parameters and structure designs of the GNN to relieve computation and communication burdens while optimizing system performance.

\vspace{-0.2em}
\begin{figure}[!h]
  \centering
  \includegraphics[width=0.5\textwidth]{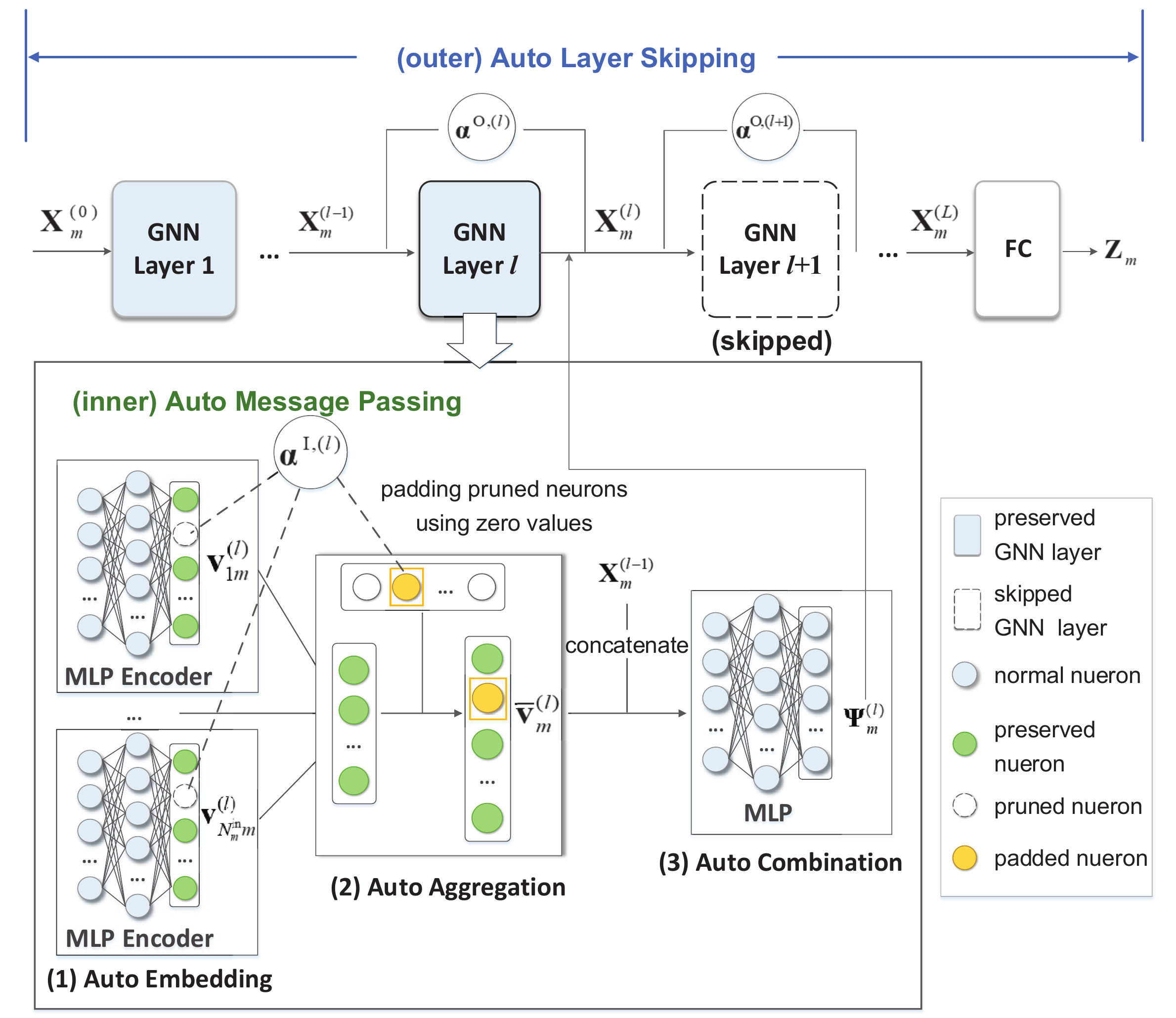}\\
  \caption{Illustration of the proposed AutoGNN architecture.}\label{fig_AutoGNN}
\end{figure}

As shown in Fig. \ref{fig_AutoGNN}, the proposed AutoGNN architecture has a dual-loop auto-learned structure, which consists of an inner auto message passing module and an outer auto layer skipping module. 
These auto-learned modules can adaptively configure the network widths (namely the message embedding sizes) of the inner MLP encoders and the network depth (namely the number of layers) of the GNN.

\subsubsection{Auto Message Passing Module}
In each GNN layer, the auto message passing module consists of three steps, namely the auto message embedding, the auto message aggregation, and the auto message combination.

\textbf{(i) Auto message embedding.} 
Define $\phi_{\mathrm{E}}^{l}$ as the local MLP encoder in GNN layer $l$. 
The high-dimensional node/edge features can be embedded by the local MLP encoder, and BSs only need to exchange the low-dimensional embedded messages for achieving coordination. 
Different from conventional GNN that employs predefined message embedding sizes, the auto message passing module enables a self-optimized message embedding size for each GNN layer,  
which can fully exploit the potential of deep learning to efficiently reduce the dimensions of the embedded messages exchanged among BSs. 

To be more specific, the auto message passing module would selectively prune unnecessary output neurons and automatically configure the network width of the MLP encoder. 
We define the binary vector $\bm{\alpha}^{\mathrm{I},(l)}=\left[\alpha_{1}^{\mathrm{I},(l)},\alpha_{2}^{\mathrm{I},(l)},...,\alpha_{D^{\mathrm{E}}}^{\mathrm{I},(l)}\right]\in\mathbb{R}^{1\times D^{\mathrm{E}}}$ to specify the selective pruning of neurons, 
where $\alpha_{i}^{\mathrm{I},(l)}=1$ if the $l$-th neuron is reserved, and $\alpha_{i}^{\mathrm{I},(l)}=0$ otherwise.
To carry out message aggregation and combination, the sizes of the adapted messages received by each agent should be consistent with the input size of the combination function $\psi_{\mathrm{C}}^{l}(\cdot)$.
Hence, upon receiving the messages, each agent would fill the pruned neurons with zero values. 
Moreover, we assume no neurons would be pruned in the first GNN layer. 
Therefore, the resulting message $\mathbf{v}_{mn}^{(l)}$ received by agent $n$ from agent $m\in\mathcal{N}_n^{\mathrm{in}}$ can be modelled by
\vspace{-0.2em}
\begin{equation}\label{LocalEmbed_AutoGNN}
\mathbf{v}_{mn}^{(l)}\!=\!\begin{cases}
\phi_{\mathrm{E}}^{l}\left(\mathbf{O}_{m}^{\mathrm{N}},\mathbf{O}_{mn}^{\mathrm{E}}\right), &\!\! l\!=\!1, \\
\!\left(\bm{\alpha}^{\mathrm{I},(l)}\right)^{T}\!\!\phi_{\mathrm{E}}^{l}\!\left(\mathbf{X}_{m}^{(l-1)},
\mathbf{O}_{mn}^{\mathrm{E}}\right)
, &\!\! l\!>\!1.
\end{cases}
\end{equation}
\vspace{-0.2em}

\textbf{(ii) Auto message aggregation.}
After collecting the flexibly embedded messages $\mathbf{v}_{nm}^{(l)}$ as given in \eqref{LocalEmbed_AutoGNN} from all neighbouring agents $m\in\mathcal{N}_{m}^{\mathrm{in}}$, 
agent $m$ aggregates the messages using a permutation-equivalent aggregation function $\phi_{\mathrm{A}}(\cdot)$.
Therefore, the aggregated features $\overline{\mathbf{v}}_{m}^{(l)}$ at each agent $m$ can be written as
\vspace{-0.2em}
\begin{equation}\label{Agg_AutoGNN}
\overline{\mathbf{v}}_{m}^{(l)}=\phi_{\mathrm{A}}\left(\left\{ \mathbf{v}_{nm}^{(l)}\right\} _{n\in \mathcal{N}_{m}^{\mathrm{in}}}\right), \forall m \in \mathcal{M}, ~ l \in \mathcal{L}.
\vspace{-0.2em}
\end{equation}

\textbf{(iii) Auto message combination.}
To update the hidden state variables at layer $l$, each agent $m$ can combine the aggregated feature $\overline{\mathbf{v}}_{m}^{l}$ with the previous hidden state  $\mathbf{X}_{m}^{(l-1)}$ through the combination function $\psi_{\mathrm{C}}^{l}(\cdot)$, as
\vspace{-0.2em}
\begin{equation}\label{Comb_AutoGNN}
\bm{\Psi}_{m}^{(l)} = \psi_{\mathrm{C}}^{l}\left(\mathbf{X}_{m}^{(l-1)},\overline{\mathbf{v}}_{m}^{(l)}\right), \forall m \in \mathcal{M}, ~ l \in \mathcal{L}.
\vspace{-0.2em}
\end{equation}

\begin{algorithm}[!t]
  \caption {Distributed Scheduling Based on the proposed AutoGNN architecture}
  \begin{algorithmic}[1]
  \REQUIRE The GNN model weights $\bm{\theta}$, architecture parameters $\bm{\alpha}$, and the channel samples.
  \FOR {each GNN layer $l\in\mathcal{L}$}
      \FOR {each BS (agent) $m\in\mathcal{M}$}
          \IF{$\alpha^{\mathrm{O},(l)}=0$}
              \STATE Skip the current GNN layer.
          \ELSE
              \STATE Each agent $m$ receives the pruned embedded messages from neighbouring agents and fill the pruned neurons with zero values, as indicated by \eqref{LocalEmbed_AutoGNN}.
              \STATE Each agent $m$ aggregate the received messages $\mathbf{v}_{nm}^{(l)}$, $\forall n\in\mathcal{N}_{m}^{\mathrm{in}}$, using \eqref{Agg_AutoGNN}, and update the local hidden state using \eqref{Comb_AutoGNN} and \eqref{HiddenState_AutoGNN}.
          \ENDIF
      \ENDFOR
  \ENDFOR
  \STATE Predict the local optimization variables at each BS $m$ using \eqref{FGNN_Output}.
  \end{algorithmic}
  \label{alg:GNN}
  \end{algorithm}

\subsubsection{Auto Layer Skipping Module}
Compared to conventional GNN architectures with fixed layers, the outer auto layer skipping module learns to adaptively skip insignificant GNN layers and avoid unnecessary message passing process to reduce 
both computation complexity and communication overheads. 

Define $\bm{\alpha}^{\mathrm{O}}=\left[\alpha^{\mathrm{O},(l)},\alpha^{\mathrm{O},(l)},...,\alpha^{\mathrm{O},(L)}\right]^{T}$ as the binary indicator of layer skipping, 
where $\alpha^{\mathrm{O},(l)} = 0$ means the $l$-th GNN layer is skipped, and $\alpha^{\mathrm{O},(l)} = 1$ otherwise. 
Then, the hidden state at each GNN layer $l$ can be updated by
\vspace{-0.2em}
\begin{equation}\label{HiddenState_AutoGNN}
\mathbf{X}_{m}^{(l)}
\!=\!\alpha^{\mathrm{O},(l)}\bm{\Psi}_{m}^{(l)}\!+\!\left(\!1\!-\!\alpha^{\mathrm{O},(l)}\!\right)\!\mathbf{X}_{m}^{(l-1)},
~ \forall m \in \mathcal{M},  l \in \mathcal{L}.
\vspace{-0.2em}
\end{equation}

Eventually, the optimization variables can be predicted by the FC layer $\varphi_{\mathrm{FC}}(\cdot)$, similar to \eqref{FGNN_Output}. 
The distributed scheduling algorithm using the proposed AutoGNN architecture can be summarized in Algorithm \ref{alg:GNN}.

\section{Bi-Level AutoGNN Learning Algorithm}
To achieve communication-efficient scheduling with the proposed AutoGNN architecture, we should jointly train GNN model weights $\bm{\theta}$ and architecture parameters $\bm{\alpha}$ to predict desirable solutions.
In this section, we first formulate the AutoGNN training as a bi-level programming, where the GNN model weights are optimized in the lower level, 
whilst the architecture parameters are optimized in the upper level.
Then, a bi-level AutoGNN learning algorithm is developed, which can efficiently calculate hypergradient for AutoGNN training.

\vspace{-0.1em}
\subsection{Bi-Level Programming for AutoGNN Learning}
Based on the proposed AutoGNN architecture, the achievable data rate of user $k\in\mathcal{K}^{m}$ in \eqref{SICConstraint} can be rewritten as
\vspace{-0.2em}
\begin{equation}\label{rate_AL}
    \begin{split}
      R_{k}^{m}\!\left(\bm{\theta},\bm{\alpha}\right)
    \!&=\!
    \! \min_{i\in\mathcal{K}^{m}}\! \bigg\{\! \beta_{ik}^{m} \! \left(\bm{\theta},\bm{\alpha}\right)r_{ik}^{m}\left(\!\bm{\theta},\!\bm{\alpha}\right)  
    \\& + \left(1\!-\!\beta_{ik}^{m}\left(\bm{\theta},\bm{\alpha}\right)\right)r_{kk}^{m}\left(\bm{\theta},\bm{\alpha}\right)\!\!\bigg\}, 
    ~ \forall k \in \mathcal{K}_{m}, 
    \end{split}
\end{equation}
\vspace{-0.4em}\\ \noindent
where $\bm{\alpha}=\left\{\bm{\alpha}^{\mathrm{O}},\bm{\alpha}^{\mathrm{I}}\right\}$ is the combined architecture parameter vector.
Then, the achievable system sum rate of the cluster-free NOMA networks can be formulated as $R\left(\bm{\theta},\bm{\alpha}\right) = \sum\limits_{m\in\mathcal{M}} \sum\limits_{k\in\mathcal{K}^{m}}R_{k}^{m}\left(\bm{\theta},\bm{\alpha}\right)$.

Let ${R}\left(\bm{\theta},\bm{\alpha}\right)$ and $\widehat{R}_{\mathrm{v}}\left(\bm{\theta},\bm{\alpha}\right)$ be the achieved sum rate during training and validation, respectively. 
The joint learning of architecture parameters $\bm{\alpha}$ and GNN model weights $\bm{\theta}$ can be formulated as a bi-level programming problem.
In the inner loop, the GNN model weights $\bm{\theta}$ are trained to maximize the training-stage sum rate ${R}\left(\bm{\theta},\bm{\alpha}\right)$ under fixed $\bm{\alpha}$.
In the outer loop, the optimal architecture parameters $\bm{\alpha}$ are searched to 
maximize the validation sum rate $\widehat{R}_{\mathrm{v}}\left(\bm{\theta},\bm{\alpha}\right)$ while fixing $\bm{\theta}$. 
This bi-level joint optimization problem can be written as
\vspace{-0.2em}
\begin{subequations}\label{P_AGNN}
\begin{align*}
& \min_{\bm{\alpha}} -\widehat{R}_{\mathrm{v}}\left(\mathbf{\bm{\theta}}^{*}\left(\bm{\alpha}\right),\bm{\alpha}\right) \tag{\ref{P_AGNN}{a}}
\\{\text{s.t.}}~ &
\mathbf{\bm{\theta}}^{*}\left(\bm{\alpha}\right)=\mathop{\arg\min}_{\bm{\theta}} -{R}\left(\bm{\theta},\bm{\alpha}\right), \label{constraint_theta} \tag{\ref{P_AGNN}{b}}
\\&
\alpha_{i}^{\mathrm{I},(l)}, \alpha^{\mathrm{O},(l)} \in \{0,1\}  ~\forall 1 \le i \le D^{\mathrm{E}}, l \in \mathcal{L}, \label{constraint_alp_bin}  \tag{\ref{P_AGNN}{c}}
\\&
\beta_{ik}^{m}\left(\bm{\theta},\bm{\alpha}\right),\zeta_{ik}^{m}\left(\bm{\theta},\bm{\alpha}\right)\in\left\{ 0,1\right\},  \label{constraint_bet_bin}  \tag{\ref{P_AGNN}{d}}
\\&
\sum\limits _{k\in\mathcal{K}_{m}}\left\Vert \mathbf{w}_{k}^{m}\left(\bm{\theta},\bm{\alpha}\right)\right\Vert ^{2}\le P^{\max}, ~\forall m\in\mathcal{M}, \label{constraint_power_AGNN} \tag{\ref{P_AGNN}e}
\\&
\beta_{ik}^{m}\left(\bm{\theta},\bm{\alpha}\right)\!+\!\beta_{ki}^{m}\left(\bm{\theta},\bm{\alpha}\right)
\!+\!\zeta_{ik}^{m}\left(\bm{\theta},\bm{\alpha}\right)\!=\!1, ~\forall i\ne k. \label{SIC_decoding_AGNN} \tag{\ref{P_AGNN}f}
\end{align*}
\end{subequations}
\vspace{-1.2em}

Before introducing the bi-level learning algorithm, we first discuss how to guarantee the constraints of the optimization problem \eqref{P_AGNN} in the learning-based algorithms. 
\subsubsection{Continuous constraint guarantees}
The transmission power constraints \eqref{constraint_power_AGNN} can be directly ensured by projecting the 
decision variable $\mathbf{W}^{m}=\mathbf{W}^{m}\left(\bm{\theta},\bm{\alpha}\right)$ output by GNN onto the feasible region, i.e.,
\vspace{-0.2em}
\begin{equation}\label{power_project}
\mathbf{W}^{m}
\!:= \!\Pi_{\mathcal{W}}\left\{ \mathbf{W}^{m}\right\}
\!\!=\!\!\begin{cases}
\mathbf{W}^{m}, &\!\!\!\text{if}\!\!\sum\limits_{k\in\mathcal{K}^{m}}\!\!\!\!\left\|\mathbf{w}_{k}^{m}\right\|^2\!\le \!P^{\max},\\
\frac{\mathbf{W}^{m}\!\sqrt{P^{\max}}\!\!}{\left\Vert \mathbf{W}^{m}\right\Vert }, & \! \text{otherwise.}
\end{cases}
\vspace{-0.2em}
\end{equation}
Furthermore, the equality constraint \eqref{SIC_decoding_AGNN} can be ensured using the softmax activation function $\mathrm{Softmax}(\cdot)$. 
Specifically, given variables $\mathbf{x}_1, \mathbf{x}_2, ..., \mathbf{x}_N$ output by the final GNN layer, 
constraint $\sum_{n=1}^{N}{\mathbf{x}_{n}} = 1$ can be stringently guaranteed  
by normalizing $\mathbf{x}_1, \mathbf{x}_2, ..., \mathbf{x}_N$ via $\mathrm{Softmax}(\cdot)$ as
\vspace{-0.2em}
\begin{equation*}
\mathrm{Softmax}\left(\left[\mathbf{x}_{1},\mathbf{x}_{2},...,\mathbf{x}_{N}\right]\right)
\!\triangleq\!\!\left[\frac{e^{\mathbf{x}_{1}}}{\sum\limits_{n}e^{\mathbf{x}_{n}}}\!,\frac{e^{\mathbf{x}_{2}}}
{\sum\limits_{n}e^{\mathbf{x}_{n}}}\!,...,\frac{e^{\mathbf{x}_{N}}}{\sum\limits_{n}e^{\mathbf{x}_{n}}}\!\right]\!.
\vspace{-0.2em}
\end{equation*}
Hence, the equality constraints \eqref{SIC_decoding_AGNN} can be enforced by directly normalizing the SIC operation variables as
\begin{equation}\label{Softmax_bet}
\left[\beta_{ik}^{m},\beta_{ki}^{m},\zeta_{ik}^{m}\right]
\!:=
\mathrm{\mathrm{Softmax}}\left(\left[{\beta}_{ik}^{m}\left(\bm{\theta}\!,\bm{\alpha}\right),{\beta}_{ki}^{m}\left(\bm{\theta},\bm{\alpha}\right)\!,
{\zeta}_{ik}^{m}\left(\bm{\theta},\bm{\alpha}\right)\right]\right)\!.
\end{equation}

\subsubsection{Binary guarantees}
To deal with the non-differentiable binary constraints \eqref{constraint_alp_bin}-\eqref{constraint_bet_bin}, we further invoke the Gumbel-Softmax reparameterization.
Given the output variable $x$,
Gumbel-Softmax randomly samples a binary variable $\widehat{x}$ according to the probabilities $P(\widehat{x}=1)=P_1 = \mathrm{Sigmoid}\left(x\right) = \frac{e^{x}}{1+e^{x}}$ 
and $P(\widehat{x}=0)=P_0=1-\mathrm{Sigmoid}\left(x\right)$. 
The sampling process of $\widehat{x}$ can be reparameterized as
\begin{equation}\label{alp_O_reparameterization}
  \widehat{x}=\mathop{\arg\max}_{j\in\{0,1\}} \left(\log P_j+u_{j}\right),
\end{equation}
where $u_{j}$, $j \in \{0,1\}$, denotes the random variables following the Gumbel distribution. 
By replacing the non-continuous function $\arg\max\left(\cdot\right)$ with the $\mathrm{Softmax}(\cdot)$ function, 
we can obtain the following differentiable binary sample 
\begin{equation}\label{alp_reparameterization}
  \widehat{x} \!= \!
  \frac{\exp\left(\frac{\log\left(P_{1}\right)\!+\!u_{1}}{s_{\mathrm{temp}}}\right)}{\sum_{j}\exp\left(\frac{\log\left(P_{j}\right)\!+\!u_{j}}{s_{\mathrm{temp}}}\right)}
  \!=\! 
  \mathrm{Sigmoid}\!\left(\frac{x\!+\!u_1\!-\!u_0}{s_\mathrm{temp}}\right),
\end{equation}
where $s_\mathrm{temp}$ is the temperature of the softmax function, which can control the discrepancy between the $\arg\max$ and the softmax functions.
Since two Gumbels' difference is a Logistic distribution, i.e., $u_1 - u_0 = \log U - \log(1-U)$ with $U\sim\mathrm{Uniform}(0,1)$, 
we can recast the sampling process as $\widehat{x} = G(x) = \mathrm{Sigmoid}\left(\frac{x+\log U - \log (1-U)}{s_\mathrm{temp}}\right)$.   
Therefore, the binary constraints \eqref{constraint_alp_bin} and \eqref{constraint_bet_bin} can be replaced by applying the following operations to decision variables output by GNN: 
\vspace{-0.2em}
\begin{equation}\label{alp_I_gumbel}
  \widehat{\alpha}_{i}^{\mathrm{I},(l)} \!=\! G\left(\alpha_{i}^{\mathrm{I},(l)}\right)\!,
  ~\forall 1\le i \le D^{\mathrm{E}}, ~ l \in \mathcal{L},
\vspace{-0.2em}
\end{equation}
\vspace{-0.2em}
\begin{equation}\label{alp_O_gumbel}
  \widehat{\alpha}^{\mathrm{O},(l)} \!=\! G\left(\alpha^{\mathrm{O},(l)}\right)\!,
  ~ \forall l \in \mathcal{L},
\vspace{-0.2em}
\end{equation}
\vspace{-0.2em}
\begin{equation}\label{alp_bet_gumbel}
  \widehat{\beta}_{i,k}^{m} \!=\! G\!\left(\beta_{i,k}^{m}\right)\!, 
  ~\widehat{\zeta}_{i,k}^{m} \!=\! G\!\left(\zeta_{i,k}^{m}\right)\!, 
  ~\forall i,k \in \mathcal{K}_{m}, ~\forall m \in \mathcal{M}.
\vspace{-0.2em}
\end{equation}

By ensuring constraints \eqref{constraint_alp_bin}  - \eqref{SIC_decoding_AGNN} in problem \eqref{P_AGNN} via the operations \eqref{power_project}, \eqref{Softmax_bet}, and \eqref{alp_I_gumbel}-\eqref{alp_bet_gumbel}, 
we can directly deal with the following unconstrained bi-level programming
\vspace{-0.2em}
\begin{subequations}\label{P_AGNN_uncon}
\begin{align*}
& \min_{\bm{\alpha}} ~ \mathcal{L}_{\mathrm{v}}\left(\bm{\theta}^{*}(\bm{\alpha}),\bm{\alpha}\right) \tag{\ref{P_AGNN_uncon}{a}}
\\{\text{s.t.}}~ &
\mathbf{\bm{\theta}}^{*}\left(\bm{\alpha}\right)=\mathop{\arg\min}_{\bm{\theta}} \mathcal{L}\left(\bm{\theta},\bm{\alpha}\right).
\label{theta_fun_AGNN}\tag{\ref{P_AGNN_uncon}{b}}
\end{align*}
\end{subequations}
\vspace{-0.5em}

\vspace{-1.3em}
\subsection{Bi-Level AutoGNN Learning Algorithm}
We develop a bi-level AutoGNN learning algorithm to handle problem \eqref{P_AGNN_uncon} in this part.
Here, we refer to the gradients of the outer-loop loss function with respect to architecture parameters $\bm{\alpha}$ as hypergradient\cite{iDARTS,HyperNeumann}.
Let $\bm{\Theta} \triangleq \bm{\theta}^{*} = \bm{\theta}^{*}\left(\bm{\alpha}\right)$ denote the optimal model weights obtained by the inner-loop optimization, which is a function of $\bm{\alpha}$ as defined in \eqref{theta_fun_AGNN}.
Based on the chain rule, the hypergradient can be derived as
\vspace{-0.2em}
\begin{equation}\label{hypergradient}
\nabla_{\bm{\alpha}}\mathcal{L}_{\mathrm{v}}=\underset{\text{direct gradient}}{\underbrace{\frac{\partial\mathcal{L}_{\mathrm{v}}\left(\bm{\Theta},\bm{\alpha}\right)}{\partial\bm{\alpha}}}}+\underset{\text{indirect gradient}}{\underbrace{\frac{\partial\mathcal{L}_{\mathrm{v}}\left(\bm{\Theta},\bm{\alpha}\right)}{\partial\bm{\Theta}}\overset{\text{\tiny{best-response Jacobian}}}{\overbrace{\nabla_{\bm{\alpha}}\bm{\theta}^{*}\left(\bm{\alpha}\right)}}}},
\vspace{-0.2em}
\end{equation}
which consists of the direct and indirect components. 
Note that the direct gradient $\frac{\partial\mathcal{L}_{\mathrm{v}}\left(\bm{\Theta},\bm{\alpha}\right)}{\partial\bm{\alpha}}$ can be directly computed. 
Hence, the main difficulty to calculate \eqref{hypergradient} lies in the computation of the indirect gradient, 
where the best-response Jacobian $\nabla_{\bm{\alpha}}\bm{\theta}^{*}\left(\bm{\alpha}\right)$ should be evaluated using the local optimum $\bm{\theta}^{*}$ of the inner-loop optimization.
For simplicity, we denote $\mathcal{L}_{\mathrm{v}}\left(\bm{\Theta},\bm{\alpha}\right)=\mathcal{L}_{\mathrm{v}}$ and $\mathcal{L}\left(\bm{\theta},\bm{\alpha}\right)=\mathcal{L}$ hereinafter.
The hypergradient calculation can be discussed as follows.

\subsubsection{Unrolling-based hypergradient}
Generally, the hypergradient defined in \eqref{hypergradient} can be computed by the reverse-mode unrolling method \cite{Reversemode_2018,Reversemode_2019}, 
which takes large enough gradient descent steps in the inner loop under given $\bm{\alpha}$ to estimate the optimal $\bm{\theta}^{*}\left(\bm{\alpha}\right)$, 
and thus compute the best-response Jacobian $\nabla_{\bm{\alpha}}\bm{\theta}^{*}\left(\bm{\alpha}\right)$  in \eqref{hypergradient}.
Specifically, given an initial point $\bm{\theta}_{0}$,  the update rule of the GNN model weights based on the gradient descent at the $t$-th inner-loop step can be written as
\vspace{-0.2em}
\begin{equation}\label{multistep_gd}
\bm{\theta}_{t}=\Phi\left(\bm{\theta}_{t-1},\bm{\alpha}\right),
\vspace{-0.2em}
\end{equation}
where $\Phi\left(\bm{\theta}_{t-1},\bm{\alpha}\right)=\bm{\theta}_{t-1}-\kappa\nabla_{\bm{\theta}}\mathcal{L}\left(\bm{\theta}_{t-1},\bm{\alpha}\right)$.
Let $T$ be the total number of inner-loop optimization iterations,
we have $\bm{\theta}^{*}\left(\bm{\alpha}\right)=\Phi\left(\bm{\theta}_{T-1},\bm{\alpha}\right)
=\Phi\left(\Phi\left(...\Phi\left(\bm{\theta}_{0},\bm{\alpha}\right)...,\bm{\alpha}\right)\right)$.
Based on the chain rule, the hypergradient can be recursively derived as
\vspace{-0.2em}
\begin{equation}\label{hypergradient_reverse}
\nabla_{\bm{\alpha}}\mathcal{L}^{\mathrm{Rev}}=\frac{\partial\mathcal{L}_{\mathrm{v}}}{\partial\bm{\alpha}}
+\frac{\mathcal{L}_{\mathrm{v}}}{\partial\bm{\Theta}}\left(\sum_{t=0}^{T}V_{t}Q_{t+1}...Q_{T}\right)\!,
\vspace{-0.2em}
\end{equation}
where $Q_{t}=\nabla_{\bm{\theta}}\Phi\left(\bm{\theta}_{t-1},\bm{\alpha}\right)$ and $V_{t}=\nabla_{\bm{\alpha}}\Phi\left(\bm{\theta}_{t-1},\bm{\alpha}\right)$.

From \eqref{hypergradient_reverse}, it is intuitive that all the intermediate GNN gradients of $T$ inner-loop steps should be recorded.
To reduce the memory cost, the truncated back propagation was proposed in \cite{Reversemode_2019}, which approximately computes the hypergradient by only storing the intermediate gradients of the last $\tau$ iterations in the inner loop ($\tau\ll T$), i.e.,
\vspace{-0.2em}
\begin{equation}
\nabla_{\bm{\alpha}}\widehat{\mathcal{L}}_{\tau}^{\mathrm{Trun}}
\!=\!\frac{\partial\mathcal{L}_{\mathrm{v}}}{\partial\bm{\alpha}}
+\frac{\partial\mathcal{L}_{\mathrm{v}}}{\partial\bm{\Theta}}\!
\left(\sum_{t=T-\tau+1}^{T}\!\!\!\!V_{t}Q_{t+1}...Q_{T}\right).
\vspace{-0.2em}
\end{equation}
However, this method still suffers unaffordable memory costs when training a large number of neural network parameters, making it impracticable and inapplicable for deep learning.
To achieve cost-efficient computation, we approximate the hypergradient using implicit function theorem (IFT) \cite{IFT_2000,HyperNeumann}, which can efficiently compute the hypergradient without recording any intermediate gradients, as analyzed as follows.

\subsubsection{Implicit hypergradient}
We invoke IFT to equivalently transform the hypergradient.
To begin with, we introduce the following assumptions for the inner and outer loss functions, which are commonly considered in the differentiable bi-level learning algorithms \cite{iDARTS,Reversemode_2018,compHyper_2020}.
\vspace{-0.2em}
\begin{assumption}\label{Assumption_L1}
The non-convex inner-loop function $\mathcal{L}\left(\bm{\theta},\bm{\alpha}\right)$ has the following properties:
\begin{enumerate}
  \item[(i)] Function $\bm{\theta}^{*}\left(\bm{\alpha}\right)$ is Lipschitz continuous with constant $L^{\alpha}>0$, and has Lipschitz-continuous gradient with constant $\widetilde{L}^{\alpha}>0$.
  \item[(ii)] Function $\mathcal{L}\left(\bm{\theta},\bm{\alpha}\right)$ is twice differentiable with respect to $\bm{\theta}$ and has Lipschitz-continuous gradient with constant $\widetilde{L}^{\theta} > 0$,
  i.e., $\left\Vert \frac{\partial\mathcal{L}}{\partial\bm{\theta}_{0}}-\frac{\partial\mathcal{L}}{\partial\bm{\theta}_{1}}\right\Vert \le \widetilde{L}^{\theta}\left\Vert \bm{\theta}_{0}-\bm{\theta}_{1}\right\Vert$.
      Moreover, for some constant $C^{\theta\alpha}>0$, $\left\Vert \frac{\partial^{2}\mathcal{L}}{\partial\bm{\alpha}\partial\bm{\theta}}\right\Vert \le C^{\theta\alpha}$.
  \item[(iii)] $\mathcal{L}\left(\bm{\theta},\bm{\alpha}\right)$ is locally strongly $\mu$-convex with respect to $\bm{\theta}$ around $\bm{\theta}^{*}\left(\bm{\alpha}\right)$, 
  meaning  that the Hessian matrix $\frac{\partial^{2}\mathcal{L}}{\partial\bm{\theta}\partial\bm{\theta}}\succeq\mu\mathbf{I}$ over a local $l_2$ ball $\mathcal{B}_{\varsigma}\left(\bm{\theta}\right):=\left\{ \bm{\theta}|\left\Vert \bm{\theta}-\bm{\theta}^{*}\left(\bm{\alpha}\right)\right\Vert \le\varsigma\left\Vert \bm{\theta}\right\Vert \right\} $ surrounding $\bm{\theta}^{*}\left(\bm{\alpha}\right)$.
\end{enumerate}
\end{assumption}
\vspace{-0.2em}

\vspace{-0.2em}
\begin{assumption}\label{Assumption_L2}
The non-convex outer-loop function $\mathcal{L}_{\mathrm{v}}\left(\bm{\Theta},\bm{\alpha}\right)$ is Lipschitz continuous with respect to $\bm{\Theta}$ and $\bm{\alpha}$ with constants $L_{\mathrm{v}}^{\Theta}>0$ and $L_{\mathrm{v}}^{\alpha}>0$, 
and has Lipschitz-continuous gradient with constants $\widetilde{L}_{\mathrm{v}}^{\Theta}>0$ and $\widetilde{L}_{\mathrm{v}}^{\alpha}>0$.
Moreover, for some constant $C_{\mathrm{v}}^{\theta}$, $\left\Vert \frac{\partial\mathcal{L}_{\mathrm{v}}}{\partial\bm{\theta}} \right\Vert \le C_{\mathrm{v}}^{\theta}$.
\end{assumption}

According to the IFT, we have the following lemma.
\vspace{-0.2em}
\begin{lemma}[\emph{Implicit Hyperegradient}]\label{lemma_ihypergradient}
Given the GNN model weights $\bm{\theta}$ that achieve the local optimum in the inner loop, i.e.,  $\frac{\partial\mathcal{L}}{\partial\bm{\theta}}\big|_{\bm{\theta}=\bm{\theta}^{*}}=0$, the hypergradient can be equivalently transformed into
\vspace{-0.2em}
\begin{equation}\label{ImplicitHypergradient}
\nabla_{\bm{\alpha}}\mathcal{L}_{\mathrm{v}}=\frac{\partial\mathcal{L}_{\mathrm{v}}}{\partial\bm{\alpha}}
-\frac{\partial\mathcal{L}_{\mathrm{v}}}{\partial\bm{\Theta}}
\mathbf{G}_{*}^{-1}\frac{\partial^{2}\mathcal{L}}{\partial\bm{\alpha}\partial\bm{\theta}}\bigg|_{\bm{\theta}=\bm{\theta}^{*}},
\vspace{-0.2em}
\end{equation}
where $\mathbf{G}_{*} = \frac{\partial^{2}\mathcal{L}} {\partial\bm{\theta}\partial\bm{\theta}}\big|_{\bm{\theta}=\bm{\theta}^{*}}$ denotes the Hessian matrix with respect to $\bm{\theta}$ at the point $\bm{\theta}^{*}$.
\begin{proof}
From $\frac{\partial\mathcal{L}}{\partial\bm{\theta}}\big|_{\bm{\theta}=\bm{\theta}^{*}}=0$, we have
\vspace{-0.2em}
\begin{equation}
\frac{\partial}{\partial\bm{\alpha}}\left(\frac{\partial\mathcal{L}}{\partial\bm{\theta}}\big|_{\bm{\theta}=\bm{\theta}^{*}}\right)=0.
\vspace{-0.2em}
\end{equation}
Therefore, we can obtain that
\vspace{-0.2em}
\begin{equation}
\frac{\partial^{2}\mathcal{L}}{\partial\bm{\alpha}\partial\bm{\theta}}\bigg|_{\bm{\theta}=\bm{\theta}^{*}}
+\frac{\partial^{2}\mathcal{L}}{\partial\bm{\theta}\partial\bm{\theta}}\frac{\partial\bm{\theta}^{*}\left(\bm{\alpha}\right)}{\partial\bm{\alpha}}\bigg|_{\bm{\theta}=\bm{\theta}^{*}}=0,
\vspace{-0.2em}
\end{equation}
which can be rearranged as
\begin{equation}\label{ImplicitHypergradient_part}
-\frac{\partial^{2}\mathcal{L}}{\partial\bm{\alpha}\partial\bm{\theta}}\bigg|_{\bm{\theta}=\bm{\theta}^{*}}
=\frac{\partial^{2}\mathcal{L}}{\partial\bm{\theta}\partial\bm{\theta}}\bigg|_{\bm{\theta}=\bm{\theta}^{*}}\frac{\partial\bm{\theta}^{*}\left(\bm{\alpha}\right)}{\partial\bm{\alpha}}.
\end{equation}
Substituting \eqref{ImplicitHypergradient_part} into \eqref{hypergradient} yields the implicit hypergradient \eqref{ImplicitHypergradient}, which completes the proof.
\end{proof}
\end{lemma}

However, it is highly computational complexity to calculate the inverse of the Hessian matrix in \eqref{ImplicitHypergradient}, especially for GNN models with massive neural parameters.
Hence, we introduce the Neumann series expansion, which provides a stable and efficient way to tractably approximate the matrix inverse.
\begin{lemma}[{Neumann series expansion \cite[Theorem~4.20]{Neumann_1998}}]\label{lemma_Neumann}
The inversion of matrix $\mathbf{G}\in\mathbb{R}^{N\times N}$ can be transformed into
\vspace{-0.2em}
\begin{equation}\label{Neumann}
\mathbf{G}^{-1}=\sum_{n=0}^{\infty}\left(\mathbf{I}-\mathbf{G}\right)^{n},
\vspace{-0.2em}
\end{equation}
if the condition $\left\Vert \mathbf{I}-\mathbf{G}\right\Vert <1$ can be satisfied, with $\Vert\cdot\Vert$ being the spectral norm.
\end{lemma}

\begin{lemma}[AutoGNN hypergradient]\label{theorem_Neumman_hypergradient}
Given a sufficiently small learning rate $\kappa$ that satisfies $\kappa<\frac{2}{\widetilde{L}^{\theta}}$, the implicit hypergradient of the proposed AutoGNN architecture can be transformed based on the Neumann series into
\vspace{-0.1em}
\begin{equation}\label{HyperGradient_Neumann}
\nabla_{\bm{\alpha}}\mathcal{L}_{\mathrm{v}}\!
=\!\frac{\partial\mathcal{L}_{\mathrm{v}}}{\partial\bm{\alpha}}\!-\!\kappa\frac{\partial\mathcal{L}_{\mathrm{v}}}{\partial\bm{\Theta}}
\left[\sum_{n=0}^{\infty}\left(\mathbf{I}-\kappa\mathbf{G}_{*}\right)^{n}\right]\frac{\partial^{2}\mathcal{L}}{\partial\bm{\alpha}\partial\bm{\theta}}\bigg|_{\bm{\theta}=\bm{\theta}^{*}}.
\end{equation}

\begin{proof}
See Appendix \ref{proof_theorem_Neumman_hypergradient}.
\end{proof}
\end{lemma}

{
\begin{algorithm}[!t]
  \caption {Bi-Level AutoGNN Learning Algorithm}
  \begin{algorithmic}[1]
  \STATE Initialize the GNN model weights $\bm{\theta}$ and the architecture parameters $\bm{\alpha}$.
  \STATE Set the iteration number as $t=0$.
  \REPEAT
      \STATE Randomly sample mini-batches of data from the training dataset $\mathcal{D}_{\mathrm{t}}$.
      \STATE Update GNN model weights $\bm{\theta}$ based on $T$-step gradient descent.
      \STATE Calculate the Hessian matrix $\frac{\partial^{2}\overline{\mathcal{L}}^{i}}{\partial\bm{\theta}\partial\bm{\theta}}$.
      \STATE Randomly sample mini-batches of data from the validation dataset $\mathcal{D}_{\mathrm{v}}$.
      \STATE Compute the stochastic approximated AutoGNN hypergradient $\nabla_{\bm{\alpha}}\widehat{\mathcal{L}}_{\mathrm{v}}$ using \eqref{HyperGradient_Neumann_Stochastic}.
      \STATE Update architecture parameters by $\bm{\alpha} \leftarrow \bm{\alpha}-\kappa\nabla_{\bm{\alpha}}\widehat{\mathcal{L}}_{\mathrm{v}}$.
  \UNTIL{converge.}
  \ENSURE The optimal GNN architecture parameters $\bm{\alpha}$ and model weights $\bm{\theta}$.
  \end{algorithmic}
  \label{alg:AutoGNN}
\end{algorithm}
\vspace{-0.2em}
}

By leveraging the first $N_{\mathrm{G}}$ terms to approximate the Hessian matrix inverse, an approximated version of the implicit hypergradient can be given by
\vspace{-0.2em}
\begin{equation}\label{HyperGradient_Neumann_J}
\nabla_{\bm{\alpha}}\widetilde{\mathcal{L}}_{\mathrm{v}}\!
=\!\frac{\partial\mathcal{L}_{\mathrm{v}}}{\partial\bm{\alpha}}\!-\!\kappa\frac{\partial\mathcal{L}_{\mathrm{v}}}{\partial\bm{\Theta}}
\left[\sum_{n=0}^{N_{\mathrm{G}}}\left(\mathbf{I}-\kappa\mathbf{G}_{*}\right)^{n}\right]\frac{\partial^{2}\mathcal{L}}{\partial\bm{\alpha}\partial\bm{\theta}}\bigg|_{\bm{\theta}=\bm{\theta}^{*}}.
\end{equation}
To deal with large-scale datasets in practice, we compute the loss functions based on mini-batches of the training and validation data samples, i.e., 
$\mathcal{L}=\frac{1}{S_{\mathrm{t}}}\sum\limits_{i=1}^{S_{\mathrm{t}}}\overline{\mathcal{L}}^{i}$ and
$\mathcal{L}_{\mathrm{v}}=\frac{1}{S_{\mathrm{v}}}\sum\limits_{j=1}^{S_{\mathrm{v}}}\overline{\mathcal{L}}_{\mathrm{v}}^{j}$,
where $S_{\mathrm{t}}$ and $S_{\mathrm{v}}$ denote the numbers of mini-batches sampled from training and validation datasets, 
and $\overline{\mathcal{L}}^{i}$ and $\overline{\mathcal{L}}_{\mathrm{v}}^{j}$ are the stochastic loss functions computed over the randomly sampled mini-batches $i$ and $j$, respectively.
In this way, we can derive the stochastic approximated AutoGNN hypergradient as
\vspace{-0.2em}
\begin{equation}\label{HyperGradient_Neumann_Stochastic}
\nabla_{\bm{\alpha}}\widehat{\mathcal{L}}_{\mathrm{v}}
\!=\!\frac{\partial\overline{\mathcal{L}}_{\mathrm{v}}^{j}}{\partial\bm{\alpha}}
\!-\!\kappa\frac{\partial\overline{\mathcal{L}}_{\mathrm{v}}^{j}}{\partial\bm{\Theta}}
\left[\sum_{n=0}^{N_{\mathrm{G}}}\left(\mathbf{I}-\kappa\mathbf{G}_{*}^{i}\right)^{n}\right]
\frac{\partial^{2}\overline{\mathcal{L}}^{i}}{\partial\bm{\alpha}\partial\bm{\theta}}\bigg|_{\bm{\theta}=\bm{\theta}^{*}},
\vspace{-0.2em}
\end{equation}
where $\mathbf{G}_{*}^{i}=\frac{\partial^{2}\overline{\mathcal{L}}^{i}}{\partial\bm{\theta}\partial\bm{\theta}}\big|_{\bm{\theta}=\bm{\theta}^{*}}$.

The bi-level AutoGNN learning algorithm based on the stochastic approximated hypergradient can be summarized in Algorithm \ref{alg:AutoGNN}.
To reduce computational complexity and enhance scalability, we assume all the BSs share the same GNN model weights and architecture parameters. 
These parameters are trained via centralized learning. After training, the copies of the GNN model/architecture parameters are distributed to each BS to implement distributed inference/scheduling.

\subsection{Theoretical Analysis}
The performance of the proposed AutoGNN regarding the permutation equivalent property, the approximation error, and the convergence can be theoretically analyzed as follows.

Let $\star\mathcal{M}$ denote the permutation operation of the node set $\mathcal{M}$, 
and $\star\mathcal{G}$ ($\star\mathbf{Z}$) denotes the permutation of graph $\mathcal{G}$ (optimization variables $\mathbf{Z}$) corresponding to $\star\mathcal{M}$.
Moreover, the solution predicted by the GNN can be written as $\mathbf{Z} = F_{\mathrm{A}}\left(\mathcal{G}\right)$, where $F_{\mathrm{A}}$ depicts the function of AutoGNN.
\begin{proposition}
The proposed AutoGNN satisfies the permutation equivalence property $\mathbf{Z}=F_{\mathrm{A}}\left(\mathcal{G}\right)$, which satisfies $F_{\mathrm{A}}\left(\star\mathcal{G}\right)=\star\left(F_{\mathrm{A}}\left(\mathcal{G}\right)\right)=\star\mathbf{Z}$.
\begin{proof}
By sharing the GNN model weights $\bm{\theta}$ and the architecture parameters $\bm{\alpha}$ among distributed agents, the proposed auto-learned module would not impact the permutation invariance property, which can be proven referring to \cite{MPGNN_2021}.
\end{proof}
\end{proposition}

The approximation error $\delta$ between the actual hypergradient $\nabla_{\bm{\alpha}}{\mathcal{L}}_{\mathrm{v}}$ with $N_{\mathrm{G}}\to\infty$
and the approximated AutoGNN hypergradient $\nabla_{\bm{\alpha}}\widetilde{\mathcal{L}}_{\mathrm{v}}$ with $N_{\mathrm{G}}<\infty$ can be defined as 
\vspace{-0.2em}
\begin{equation}\label{appr_error}
\delta\triangleq \nabla_{\bm{\alpha}}\mathcal{L}_{\mathrm{v}}-\nabla_{\bm{\alpha}}\widetilde{\mathcal{L}}_{\mathrm{v}},
\vspace{-0.2em}
\end{equation}
which can be bounded according to the following Lemma.
\begin{lemma}\label{lemma_error}
Under Assumption \ref{Assumption_L1}, error $\delta$ is upper bounded by
\vspace{-0.2em}
\begin{equation}\label{error_hypergradient}
\left\Vert\delta\right\Vert \le C^{\theta\alpha}C_{\mathrm{v}}^{\theta}\frac{1}{\mu}\left(1-\kappa\mu\right)^{N_{\mathrm{G}}+1}.
\vspace{-0.2em}
\end{equation}
\begin{proof}
Based on the definitions of $\nabla_{\bm{\alpha}}\widetilde{\mathcal{L}}_{\mathrm{v}}$ and $\nabla_{\bm{\alpha}}\mathcal{L}_{\mathrm{v}}$, we have
\vspace{-0.2em}
\begin{equation}\label{bound_L-tildeL}
\nabla_{\bm{\alpha}}\mathcal{L}_{\mathrm{v}}\!-\!\nabla_{\bm{\alpha}}\widetilde{\mathcal{L}}_{\mathrm{v}}
\!=\!\kappa\frac{\partial\mathcal{L}_{\mathrm{v}}}{\partial\bm{\theta}}
\sum_{n=N_{\mathrm{G}}+1}^{\infty}\!\!\!\left[\mathbf{I}-\kappa\mathbf{G}_{*}\right]^{n}\frac{\partial^{2}\mathcal{L}}{\partial\bm{\alpha}\partial\bm{\theta}}.
\vspace{-0.2em}
\end{equation}
Since function $\mathcal{L}$ is locally $\mu$-strongly convex and has Lipschitz-continuous gradient surrounding $\bm{\theta}^{*}$, we have $\kappa\mu\mathbf{I}\preceq\kappa\mathbf{G}_{*}\preceq\mathbf{I}$ with $\kappa<L^{\theta}$, which yields
\vspace{-0.2em}
\begin{equation}\label{bound_I-kG}
\sum_{n=N_{\mathrm{G}}+1}^{\infty}\!\!\!\!\!\left[\mathbf{I}\!-\!\kappa\mathbf{G}_{*}\right]^{n}
\!\le\!\!\!\sum_{n=N_{\mathrm{G}}+1}^{\infty}\!\!\!\!\left[1\!-\!\kappa\mu\right]^{n}
\!\overset{(a)}{\le}\!\frac{1}{\kappa\mu}\!\left(1\!-\!\kappa\mu\right)^{N_{\mathrm{G}}\!+\!1}\!,
\vspace{-0.2em}
\end{equation}
where $(a)$ is obtained using the sum rate of the geometry sequence.
Considering $\left\Vert \frac{\partial\mathcal{L}_{\mathrm{v}}}{\partial\bm{\theta}}\right\Vert \le C_{\mathrm{v}}^{\theta}$ and $\left\Vert \frac{\partial^{2}\mathcal{L}}{\partial\bm{\alpha}\partial\bm{\theta}}\right\Vert \le C^{\theta\alpha}$  and substituting \eqref{bound_I-kG} into \eqref{bound_L-tildeL}, we have
\vspace{-0.2em}
\begin{equation}
\left\Vert \nabla_{\bm{\alpha}}\mathcal{L}_{\mathrm{v}}-\nabla_{\bm{\alpha}}\widetilde{\mathcal{L}}_{\mathrm{v}}\right\Vert
\le C_{\mathrm{v}}^{\theta}C^{\theta\alpha}\frac{1}{\mu}\left(1-\kappa\mu\right)^{N_{\mathrm{G}}+1},
\vspace{-0.2em}
\end{equation}
which ends the proof.
\end{proof}
\end{lemma}

The convergence performance of the proposed AutoGNN can be characterized by the following theorem.
\vspace{-0.2em}
\begin{theorem}\label{theorem_convergence}
Under Assumption \ref{Assumption_L1} - \ref{Assumption_L2},
the proposed AutoGNN algorithm based on the stochastic approximated hypergradient can converge to a stationary point when the learning rate $\kappa$ is sufficiently small, namely,
\vspace{-0.2em}
\begin{equation}\label{convergence}
\lim_{u\to\infty}\mathbb{E}\left[\left\Vert \nabla_{\bm{\alpha}}\widehat{\mathcal{L}}_{\mathrm{v}}^{i}\left(\bm{\Theta}^{u},\bm{\alpha}^{u}\right)\right\Vert \right]=0, 
\vspace{-0.2em}
\end{equation}
where $\bm{\Theta}^{u}$ and $\bm{\alpha}^{u}$ denote the GNN model weights and architecture parameters at the outer-loop iteration $u$.
\begin{proof}
See Appendix \ref{proof_theorem_convergence}.
\end{proof}
\end{theorem}
\vspace{-1em}

\section{Simulation Results}
In this section, we introduce several benchmark algorithms based on conventional optimization-based algorithms, conventional fixed GNN, and CNN. 
Then, we present numerical results to verify the effectiveness of the proposed algorithm.

\vspace{-0.2em}
\subsection{Benchmark Algorithms}
\subsubsection{Distributed ADMM design}\label{Sec_Dis_ADMM} 
A general distributed optimization method to deal with non-convex MINLPs is distributed ADMM \cite{DecADMM_2011}, which can achieve locally optimal solutions by only exchanging some intermediate information among BSs.
Here, we develop a benchmark distributed ADMM algorithm for multi-cell cluster-free NOMA scheduling. 
To deal with binary variable $\bm{\beta}^{m}$, we first introduce the auxiliary variable $\widetilde{\bm{\beta}}^{m}=\left\{\widetilde{\beta}_{ik}^{m}\right\}$, which satisfies
\vspace{-0.2em}
\begin{equation}\label{constraint_bin1}
\bm{\beta}^{m} + \widetilde{\bm{\beta}}^{m} = \mathbf{1}_{K\times K}, ~\forall m \in\mathcal{M},
\vspace{-0.1em}
\end{equation}
\begin{equation}\label{constraint_bin0}
\beta_{ik}^{m}\widetilde{\beta}_{ik}^{m} = 0, ~\forall i,k \in\mathcal{K}_m, ~\forall m \in\mathcal{M},
\vspace{-0.1em}
\end{equation}
\begin{equation}\label{constraint_var3}
0 \le \beta_{ik}^{m} \le 1, ~\forall i,k \in\mathcal{K}_m, ~\forall m \in\mathcal{M}.
\vspace{-0.1em}
\end{equation}
Since constraints \eqref{constraint_bin1}-\eqref{constraint_var3} ensure $\beta_{ik}^{m}\left(1-\beta_{ik}^{m}\right)=0$, the original binary constraint \eqref{constraint_bin} can be equivalently replaced.
To deal with the max-min problem, we further introduce the slack variable $\bm{\Gamma} = \left\{ \Gamma_{k}^{m}\right\}$, which can be defined as
$\Gamma_{k}^{m}= \min_{i\in\mathcal{K}_{m}} \left\{ \frac{1}{\beta_{ik}}r_{ik}^{m} \right\}
\le \frac{1}{\beta_{ik}^{m}}r_{ik}^{m}, ~\forall k\in\mathcal{K}_{m}, ~ m\in\mathcal{M}$.
Then, $\mathcal{P}_{0}$ can be equivalently transferred into
\vspace{-0.1em}
\begin{subequations}\label{P1}
\begin{align*}
\mathcal{P}_{1}: ~ & \max_{\bm{\Gamma},\bm{\beta},\bm{\widetilde{\beta}},\mathbf{W}} ~
\sum\limits_{m\in\mathcal{M}} \sum\limits_{k\in\mathcal{K}_{m}}
\Gamma_{k}^{m} \tag{\ref{P1}{a}}
\\ {\mathrm{s.t.}}~~ &
\beta_{ik}^{m}\Gamma_{k}^{m} \le r_{ik}^{m},~\forall i,k \in \mathcal{K}_{m}, ~ \forall m \in\mathcal{M},  \label{constraint_X2}\tag{\ref{P1}b}
\\&
\Gamma_{k}^{m} \ge r_{k}^{m,\min},~\forall k \in \mathcal{K}_{m}, ~ \forall m \in\mathcal{M},  \label{constraint_rate2}\tag{\ref{P1}c}
\\&
\eqref{constraint_power}-\eqref{constraint_Decoding}, \eqref{constraint_bin1}-\eqref{constraint_var3}. \tag{\ref{P1}d}
\end{align*}
\end{subequations}
\vspace{-1em}

To solve $\mathcal{P}_{1}$, we employ the MMSE \cite{MMSE_Dai} to transform the non-convex data rate expression.
Based on the MMSE detection, the decoding rate can be written as
\vspace{-0.2em}
\begin{equation}\label{rate_MMSE}
r_{ik}^{m}\!=\!\max_{c_{ik}^{m}}\max_{a_{ik}^{m}>0}\!\left(\!\log_{2}a_{ik}^{m}\!-\!\frac{a_{ik}^{m}\epsilon_{ik}^{m}}{\ln2}
\!+\!\frac{1}{\ln2}\!\right)\!\!, 
~\forall i,k\in\mathcal{K}_{m},
\vspace{-0.2em}
\end{equation}
where the mean square error (MSE) $\epsilon_{ik}^{m}$ can be given by
\vspace{-0.2em}
\begin{equation*}
\epsilon_{ik}^{m} 
\!=\!1\!-\!2\mathrm{Re}\left(c_{ik}^{m}\mathbf{h}_{mi}^{m}\mathbf{w}_{u}^{m}\right)
\!+\!\left|c_{ik}^{m}\right|^{2}\left(\!\left|\mathbf{h}_{mi}^{m}\mathbf{w}_{k}^{m}\right|^{2}
\!+\!\mathrm{Intf}_{ik}^{m}\!+\!\sigma^2\!\right)\!,
\vspace{-0.2em}
\end{equation*}
with $c_{ik}^{m}$ being the channel equalization coefficient.
At each iteration, given the optimal solutions of \eqref{rate_MMSE}, i.e.,  
$c_{ik}^{m}=\frac{\left(\mathbf{h}_{mi}^{m}\mathbf{w}_{k}^{m}\right)^{H}}{\left|\mathbf{h}_{mi}^{m}\mathbf{w}_{k}^{m}\right|^{2}+\mathrm{Intf}_{ik}^{m}+\sigma^{2}}$ 
and 
$a_{ik}^{m}=\frac{\left|\mathbf{h}_{mi}^{m}\mathbf{w}_{k}^{m}\right|^{2}+\mathrm{Intf}_{ik}^{m}+\sigma^{2}}{\mathrm{Intf}_{ik}^{m}+\sigma^{2}}$ \cite{MMSE_Dai}, 
constraint (\ref{P1}b) can be recast as
\vspace{-0.2em}
\begin{equation}
\beta_{ik}^{m}\Gamma_{k}^{m}\!\le\!\log_{2}a_{ik}^{m}\!-\!\frac{a_{ik}^{m}\epsilon_{ik}^{m}}{\ln2} \!+\!\frac{1}{\ln2},
~\forall i,k\in\mathcal{K}_{m}, ~\forall m \in \mathcal{M}. 
\vspace{-0.2em}
\end{equation}
On the other hand, to deal with the highly coupling variables $\beta_{iu}^{m}$, $\beta_{uk}^{m}$, and $\beta_{ku}^{m}$ in $\mathrm{Intf}_{ik}^{m}\left(\bm{\beta}^{m},\mathbf{W}\right)$, we can rearrange
\eqref{Intf_decoding1} as
\vspace{-0.2em}
\begin{equation}\label{Intf_decoding1_convex}
\begin{split}
  &\mathrm{Intf}_{ik}^{m}\left(\widetilde{\bm{\beta}}^{m},\mathbf{W}\right)\!=\underset{\text{intra-cell interference from weaker users}}{\underbrace{\sum\limits _{u<k}\max\left\{ \!\widetilde{\beta}_{iu}^{m},1-\widetilde{\beta}_{uk}^{m}\right\} \left|\mathbf{h}_{mi}^{m}\mathbf{w}_{u}^{m}\right|^{2}}}\!
\\&+\!\underset{\text{intra-cell interference from stronger users}}{\underbrace{\sum\limits _{u>k}\max\left\{ \!\widetilde{\beta}_{iu}^{m},\!\widetilde{\beta}_{ku}^{m}\right\} \left|\mathbf{h}_{mi}^{m}\mathbf{w}_{u}^{m}\right|^{2}}}
+\mathrm{ICI}_{i}^{m}\left(\mathbf{W}\right),
\end{split}
\end{equation}
\vspace{-0.4em}\\ \noindent
Since $\max\{f(x),g(x)\}$ is convex when both functions $f(x)$ and $g(x)$ are convex, $\mathrm{Intf}_{ik}^{m}\left(\widetilde{\bm{\beta}}^{m},\mathbf{W}\right)$ in  \eqref{Intf_decoding1_convex} is convex over $\widetilde{\bm{\beta}}$.
Thereafter, problem $\mathcal{P}_{1}$ can be transferred into a multi-convex problem over $\bm{\Gamma}$, $\bm{\beta}$, $\widetilde{\bm{\beta}}$, and $\mathbf{W}$.
However, this multi-convex problem still cannot be directly decomposed among the distributed BSs owing to the ICI terms.
To decouple ICI, we introduce a slack variable $\xi_{mnk}$ that indicates the upper bound of ICI caused by BS $m$ to user $k$ served by BS $n$.
Then, each BS $m$ would store the local copies $\xi_{mnk}^{m}$ and $\xi_{nmk}^{m}$, which respectively correspond to ICI terms $\xi_{mnk}$ and $\xi_{nmk}$ that are related to BS $m$.
Based on the above definitions, the local variable $\xi_{mnk}^{m}$ at each BS $m$ satisfies
\vspace{-0.2em}
\begin{equation}\label{constraint_ICI}
\xi_{mnk}^{m}\ge \sum_{u\in\mathcal{K}_{m}} \left|\mathbf{h}_{nk}^{m}\mathbf{w}_{u}^{m}\right|^2 , ~ \forall n\ne m, ~k \in \mathcal{K}_{n}, ~ m\in\mathcal{M}.
\vspace{-0.4em}
\end{equation}

Let $\bm{\xi}^{m} = \big[\xi_{m11}^{m},\xi_{m12}^{m},...,{\xi}_{mMK}^{m}, ~ {\xi}_{1m1}^{m},{\xi}_{1m2}^{m},...,{\xi}_{MmK}^{m} \big]$ 
denote the stacked local ICI variables at BS $m$. 
Moreover, $\widehat{\xi}_{mnk}^{m}$ denotes the global copy of $\xi_{mnk}$. 
Thereafter, the consensus between distributed BSs can be achieved by forcing the local copy ${\bm{\xi}}^{m}$ and the global copy $\widehat{\bm{\xi}}^{m}$ to be equal, i.e.,
\vspace{-0.2em}
\begin{equation}\label{constraint_consensus}
\widehat{\bm{\xi}}^{m} = \bm{\xi}^{m}, ~ \forall m \in \mathcal{M}.  
\vspace{-0.2em}
\end{equation}
Therefore, we can equivalently transform $\mathcal{P}_{1}$ into
\vspace{-0.2em}
\begin{subequations}\label{PDec}
\begin{align*}
\mathcal{P}_{2}: & \max_{\bm{\Gamma},\mathbf{W},\bm{\beta},\widetilde{\bm{\beta}},\bm{\xi},\widehat{\bm{\xi}}} ~  \sum\limits_{m\in\mathcal{M}} \sum\limits_{k\in\mathcal{K}^{m}}\Gamma_{k}^{m} \tag{\ref{PDec}{a}}
\\ {\mathrm{s.t.}}~ &
\beta_{ik}^{m}\Gamma_{k}^{m}
\!\le\!\log_{2}{a}_{ik}^{m}\!-\!\frac{{a}_{ik}^{m}\widetilde{\epsilon}_{ik}^{m}}{\ln2}\!+\!\frac{1}{\ln 2},
~\forall i,k\in\mathcal{K}_{m}, \label{constraint_SIC3} \tag{\ref{PDec}{b}}
\\&
\eqref{constraint_power} - \eqref{constraint_Decoding}, \eqref{constraint_bin1} - \eqref{constraint_var3}, \eqref{constraint_rate2}, 
\eqref{constraint_ICI}-\eqref{constraint_consensus},  \tag{\ref{PDec}c}
\end{align*}
\end{subequations}

\vspace{-0.2em}
\noindent where $\widetilde{\epsilon}_{ik}^{m}$ is the local copy of $\epsilon_{ik}^{m}$ at BS $m$ with 
${\mathrm{ICI}}_{i}^{m}$ in \eqref{Intf_decoding1_convex} being replaced by $\widetilde{\mathrm{ICI}}_{i}^{m} = \sum\limits _{n\ne m}\xi_{nmi}^{m}$.
Based on ADMM, the augmented Lagrangian of $\mathcal{P}_{2}$ can be formulated as
\vspace{-0.2em}
\begin{equation*}\label{PAL}
\begin{split}
&\mathcal{L}_{\mathrm{A}}\left(\bm{\omega},\widehat{\bm{\xi}};\bm{\lambda},\widetilde{\bm{\lambda}},\bm{\nu}\right)
=\sum_{m}\mathcal{L}_{\mathrm{A}}^{m}\left(\bm{\omega}^{m},\widehat{\bm{\xi}}^{m};\bm{\lambda}^{m},\widetilde{\bm{\lambda}}^{m},\bm{\nu}^{m}\right)
\\& = \sum_{m}\bigg[\sum_{k\in\mathcal{K}_{m}}\Gamma_{k}^{m} 
-\frac{1}{2\rho}\left\Vert \bm{\beta}^{m}+\widetilde{\bm{\beta}}^{m}
-\!\mathbf{1}_{K\times K}\!+\!\rho\bm{\lambda}^{m}\right\Vert^{2} \!
\\& -\!\frac{1}{2\rho}\sum\limits_{m\in\mathcal{M}}\sum\limits_{i,k\in\mathcal{K}_{m}}\!\! \left(\beta_{ik}^{m}\widetilde{\beta}_{ik}^{m}\!
+\!\rho\widetilde{\lambda}_{ik}^{m}\right)^{2}
\!\!-\!\frac{1}{2\rho}\left\Vert \bm{\xi}^{m}\!\!-\!\!\widehat{\bm{\xi}}^{m}\!+\!\rho\bm{\nu}^{m}\right\Vert^{2}\bigg],
\end{split}
\end{equation*}
\vspace{-0.2cm} \\ \noindent
where $\bm{\lambda}$, $\widetilde{\bm{\lambda}}$, and $\bm{\nu}$ represent the dual variables corresponding to equality constraints \eqref{constraint_bin1}, \eqref{constraint_bin0}, and \eqref{constraint_ICI},  respectively.
$\bm{\omega}^{m} = \left\{ \bm{\Gamma}^{m},\mathbf{W}^{m},\bm{\beta}^{m},\widetilde{\bm{\beta}}^{m},\bm{\xi}^{m}, \widetilde{\mathbf{a}}^{m},\widetilde{\mathbf{c}}^{m}\right\}$ stacks the variables locally solved at BS $m$.
During each iteration, the distributed ADMM alternately updates the global variables $\widehat{\bm{\xi}}$, local variables $\bm{\omega}$, and dual variables $\left\{\bm{\lambda},\bm{\widetilde{\lambda}},\bm{\nu}\right\}$ as follows.

\textbf{(i) Global variable update.}
At each iteration, 
the global variables $\widehat{\bm{\xi}}_{mn}=\left[\widehat{\xi}_{mn1},\widehat{\xi}_{mn2},...,\widehat{\xi}_{mnK}\right]$ can be updated by 
solving the unconstrained convex quadratic programming $
\widehat{\bm{\xi}}_{mn}= \mathop{\arg\min}_{\bm{\widehat{\xi}}_{mn}}~\frac{1}{2\rho}\big(\bm{\xi}_{mn}^{m}-\bm{\widehat{\xi}}_{mn}+\rho\bm{\nu}_{mn}^{m}\big)^{2}
+
\frac{1}{2\rho}\big(\bm{\xi}_{mn}^{n}-\bm{\widehat{\xi}}_{mn}+\rho\bm{\nu}_{mn}^{n}\big)^{2}$, 
whose solution can be derived as
\vspace{-0.2em}
\begin{equation}\label{Sol_Intf}
\widehat{\bm{\xi}}_{mn} \leftarrow \frac{1}{2}\left[\bm{\xi}_{mn}^{m}+\bm{\xi}_{mn}^{n}+\rho\left(\bm{\nu}_{mn}^{m}+\bm{\nu}_{mn}^{n}\right)\right].
\vspace{-0.2em}
\end{equation}
From \eqref{Sol_Intf}, we can observe that only the information  
$\big[\bm{\xi}_{mn}^{m}+\frac{1}{\rho}\bm{\nu}_{mn}^{m},\bm{\xi}_{nm}^{m}+\frac{1}{\rho}\bm{\nu}_{nm}^{m}\big]$ 
and $\big[\bm{\xi}_{mn}^{n}+\frac{1}{\rho}\bm{\nu}_{mn}^{n},\bm{\nu}_{mn}^{n}+\frac{1}{\rho}\bm{\nu}_{nm}^{n}\big]$ 
should be exchanged between BS $m$ and BS $n$ to update $\bm{\widehat{\xi}}_{mn}$ and $\bm{\widehat{\xi}}_{nm}$ during each iteration.

\textbf{(ii) Local variable update.}
Given $\widehat{\bm{\xi}}^{m}$ and dual variables $\big\{\bm{\lambda}^{m},\widetilde{\bm{\lambda}}^{m},\bm{\nu}^{m}\big\}$,
each BS $m$ locally solves the decomposed variables $\bm{\omega}^{m}$ in a distributed and parallel way.
Define $\widetilde{\mathrm{Inft}}_{ik}^{m}$ as the interference calculated by $\widetilde{\mathrm{ICI}}_{ik}^{m}$ in \eqref{Intf_decoding1_convex}. 
$\widetilde{a}_{ik}^{m}$ and $\widetilde{c}_{ik}^{m}$ can be respectively updated by
\vspace{-0.2em}
\begin{equation*}
\widetilde{a}_{ik}^{m} \leftarrow 
\left({\left|\mathbf{h}_{mi}^{m}\mathbf{w}_{k}^{m}\right|^{2}
\!+\!\widetilde{\mathrm{Intf}}_{ik}^{m}\!+\!\sigma^{2}}\!\right)
\!\left({\widetilde{\mathrm{Intf}}_{ik}^{m}\!+\!\sigma^{2}}\!\right)^{\!-1}\!,
\vspace{-0.2em}
\end{equation*}
\begin{equation*}
\widetilde{c}_{ik}^{m}
\leftarrow {\left(\mathbf{h}_{mi}^{m}\mathbf{w}_{k}^{m}\right)^{H}}
\!\left({\left|\mathbf{h}_{mi}^{m}\mathbf{w}_{k}^{m}\right|^{2}+\widetilde{\mathrm{Intf}}_{ik}^{m}+\sigma^{2}}\right)^{-1}\!.
\vspace{-0.2em}
\end{equation*}
Thereafter, 
the local optimization variables can be updated by 
\vspace{-0.2em}
\begin{equation*}
  \{\bm{\xi}^{m}\! ,\bm{\beta}^{m}\! \} \leftarrow \mathop{\arg\min}_{\bm{\xi}^{m}, \bm{\beta}^{m}\in\Omega^{m}} 
  \!\mathcal{L}_{\mathrm{A}}^{m}\! \big(\bm{\omega}^{m}\!,\widehat{\bm{\xi}}^{m};
\!\bm{\lambda}^{m},\!\widetilde{\bm{\lambda}}^{m},\!\bm{\nu}^{m}\big),
\vspace{-0.2em}
\end{equation*}
\vspace{-0.2em}
\begin{equation*}
  \{\bm{\Gamma}^{m}\!,\mathbf{W}^{m}\!,\widetilde{\bm{\beta}}^{m}\!\} \leftarrow
  \mathop{\arg\min}_{\bm{\Gamma}^{m}\!,\mathbf{W}^{m}\!,\widetilde{\bm{\beta}\!}^{m}\in\Omega^{m}} 
\mathcal{L}_{\mathrm{A}}^{m} \big(\bm{\omega}^{m},\widehat{\bm{\xi}}^{m}; 
\bm{\lambda}^{m},\widetilde{\bm{\lambda}}^{m},\bm{\nu}^{m}\big),
\vspace{-0.2em}
\end{equation*}
where $\Omega^{m} = \big\{ \bm{\omega}^{m} \big| \eqref{constraint_power} - \eqref{constraint_Decoding},\eqref{constraint_var3}, \eqref{constraint_ICI}, \eqref{constraint_SIC3} \big\}$ 
denotes the local feasible set of $\bm{\omega}_{m}$. 
Moreover, the dual variables can be updated by 
$\bm{\lambda}^{m}\leftarrow\bm{\lambda}^{m}+\frac{1}{\rho}\big(\bm{\beta}^{m}+\widetilde{\bm{\beta}}^{m}-\mathbf{1}_{K\times K}\big)$,
$\widetilde{\lambda}_{ik}^{m}\leftarrow\widetilde{\lambda}_{ik}^{m}+\frac{1}{\rho}\widetilde{\beta}_{ik}^{m}\beta_{ik}^{m}$, 
and $\bm{\nu}^{m}\leftarrow {\bm{\nu}}^{m}
+\frac{1}{\rho}\big(\bm{\xi}^{m}-\widehat{\bm{\xi}}^{m}\big)$.

\subsubsection{Benchmark algorithms}
We consider four benchmark algorithms for the joint coordinated beamforming and SIC optimization in multi-cell cluster-free NOMA networks:
\begin{itemize}
  \item \textbf{Distributed ADMM}: where BSs exchange information during each iteration to achieve distributed optimization without directly sharing CSI, 
  as described in Section \ref{Sec_Dis_ADMM}.
  \item \textbf{Centralized ADMM}: where BSs directly send their local CSI to a centralized controller, 
  and the multi-cell coordinated beamforming and SIC operations are obtained by solving $\mathcal{P}_{1}$ using the MMSE reformulation and the centralized ADMM method. 
  \item \textbf{Centralized CNN}: where the centralized CNN built on VGG-16 \cite{VGG} is employed, 
  which collects the local CSI from all BSs as inputs and outputs the joint optimization variables. 
  \item \textbf{Fixed GNN}: where the message passing GNN \cite{MPGNN_2021} with the fixed architecture is employed, 
  i.e., fixed network depth $L^{F}\in\{2,4\}$ and fixed message embedding sizes $D_{F}=\{16,24,48\}$, as described in Section \ref{Sec_MPGNN}.
\end{itemize}

\vspace{-0.2em}
\subsection{Numerical Results}
We consider $M=\{3,5\}$ coordinated BSs, where each BS equips $N_{\mathrm{T}}=\{4,6\}$ antennas to serve $K=\{6,10\}$ users. The signal-to-noise-ratio (SNR) is $20$ dB. 
Users' minimum rate requirements are $R_{k}^{m, \min}=0.3$ bps/Hz, $\forall k$.
According to the exponentially correlated Rayleigh fading channel \cite{CorrChannel}, we model 
$\mathbf{H}_{mn}=\widetilde{\mathbf{H}}_{mn}\left(\mathbf{R}_{mn}^{\mathrm{C}}\right)^{1/2}$, 
where $\widetilde{\mathbf{H}}_{mn}$ represents the normalized Rayleigh fading channel and 
$\mathbf{R}_{mn}^{\mathrm{C}}\in\mathbb{C}^{K\times K}$ indicates users' spatial channel correlations. 
$\mathbf{R}_{mn}^{\mathrm{C}}$ can be modelled by a Hermitian matrix, where the $(i,k)$-th element $\mathbf{R}_{mn}^{\mathrm{C}}[i,k]=(corr\times e^{j\phi})^{k-i},~ \forall k \ge i$, 
with $corr$ controlling the mean channel correlations and $\phi \in [0,2\pi]$ being the random phase. 
We set $corr_{D}=\{0.5,0.55,0.6,0.65,0.7,0.75,0.8\}$ for data channels $\mathbf{H}_{mm}$, $\forall m$, and $corr_{I}=0.5$ for interference channels $\mathbf{H}_{mn}$, $\forall m\ne n$, respectively. 
The path loss model is $loss=\left(1+\frac{d}{d_0}\right)^{-\alpha}$ with the coefficient $\alpha=3$. 
For information exchange among BSs, we assume each real floating point number takes up $32$ bits, and each complex point number occupies $64$ bits.

Due to the high computation complexity required by the MINLP problem optimization, it is impractical to label the large-scale training dataset with high-quality solutions. 
Hence, 
we train both the fixed GNN and AutoGNN in an off-line unsupervised learning manner by minimizing the loss function based on stochastic gradient descent (SGD) with $100$ epochs. 
The training dataset, validation dataset, and the test dataset include $40$, $10$, and $10$ mini-batches, respectively.  
The batch size is $32$. 
Moreover, the training dataset and the validation dataset are randomly renewed at the beginning of each epoch to ensure data diversity.

\begin{table*}[!htbp]
  \vspace{-1em}
  \begin{minipage}{1\linewidth}
      \centering
      \caption{Comparisons of cmputation/communication overheads for different algorithms, $M=3$, $N_{\mathrm{T}}=4$, $K=6$, $corr_{D}=0.6$.} \vspace{-0.2em}
      \label{table:comp}
      \resizebox{1\textwidth}{!}{
      \begin{tabular}{l|c|c|c|c}
          \hline
             Method & Test sum rate (bps/Hz) & Execution time & Number of GNN layers/iterations & Information overhead (Kbit) \\ \hline
             Centralized ADMM &	16.17 &	12.41 min/sample & 21.45	& 21.31 \\ \hline
             Distributed ADMM	& 15.02	& 5.86 min/sample	& 12.63 & 29.09 \\ \hline
            Centralized CNN & 11.12  & 4.08 s/batch & 16 & 21.31 \\ \hline
            Fixed GNN ($L=4$,$D_{F}^{E}=48$) & 17.48  & 3.78 s/batch & 4 & 36.86\\ \hline
            AutoGEL & \textbf{18.41}  & 4.12 s/batch & 4 & 49.15 \\ \hline
            AutoGNN & 18.37	& \textbf{3.26 s/batch}	& 3	& \textbf{17.66} \\ \hline

      \end{tabular}
      }
\end{minipage}
\hfill
\vspace{-1em}
\end{table*}

\begin{figure}[!h]
  \vspace{-1.5em}
  \centering
  \includegraphics[width=0.5\textwidth]{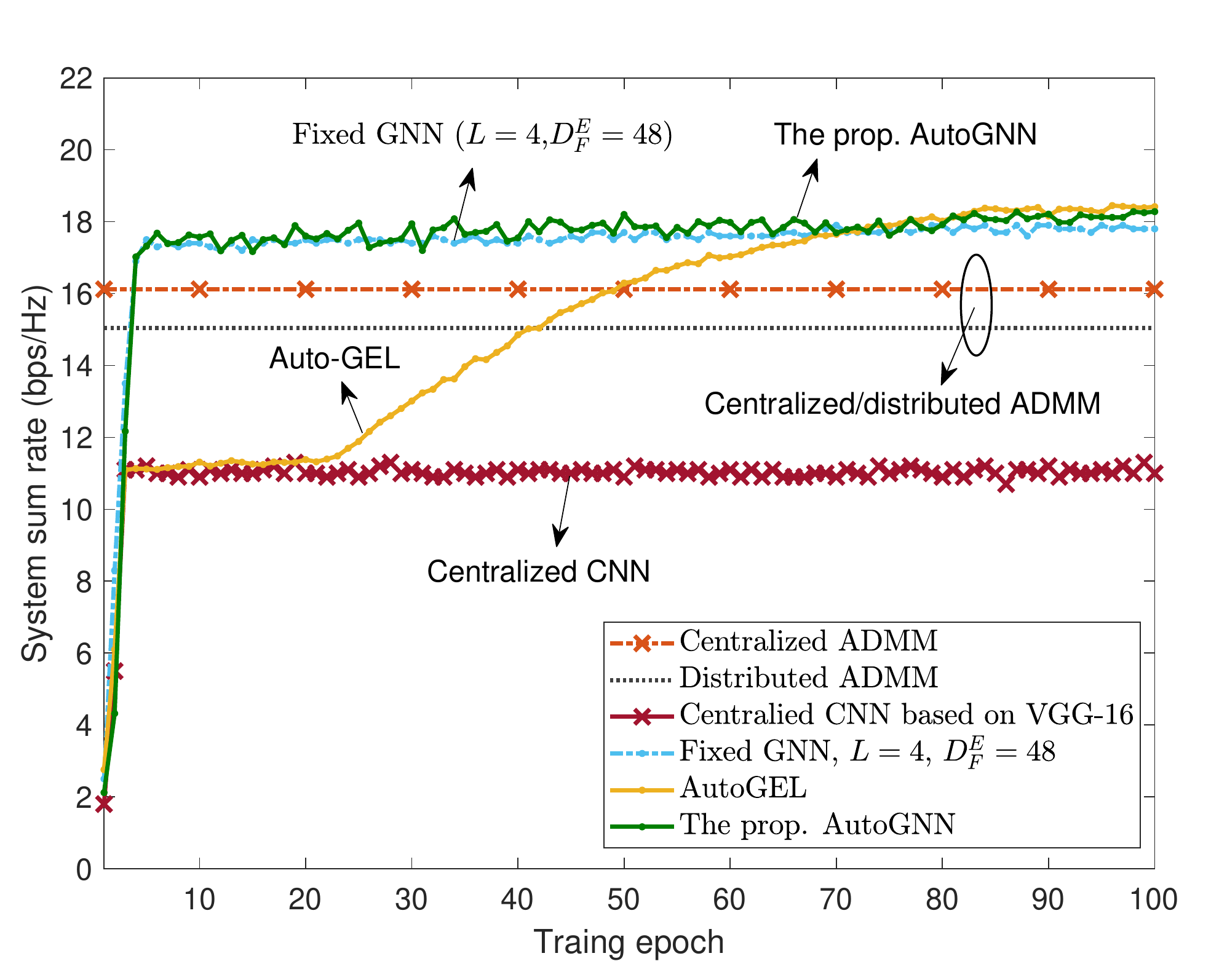}\\
  \caption{Convergence comparisons among the proposed algorithm and the benchmarks. $M=3$, $N_{\mathrm{T}}=4$, $K=6$, $corr_{D}=0.6$.}\label{fig_conv}
\end{figure}

\begin{figure}[!htbp]
  \vspace{-1.5em}
    \centering
    \subfloat[System sum rate. $M=3$, $N_{\mathrm{T}}=4$, $K=6$.]{\centering \scalebox{0.4}{\includegraphics{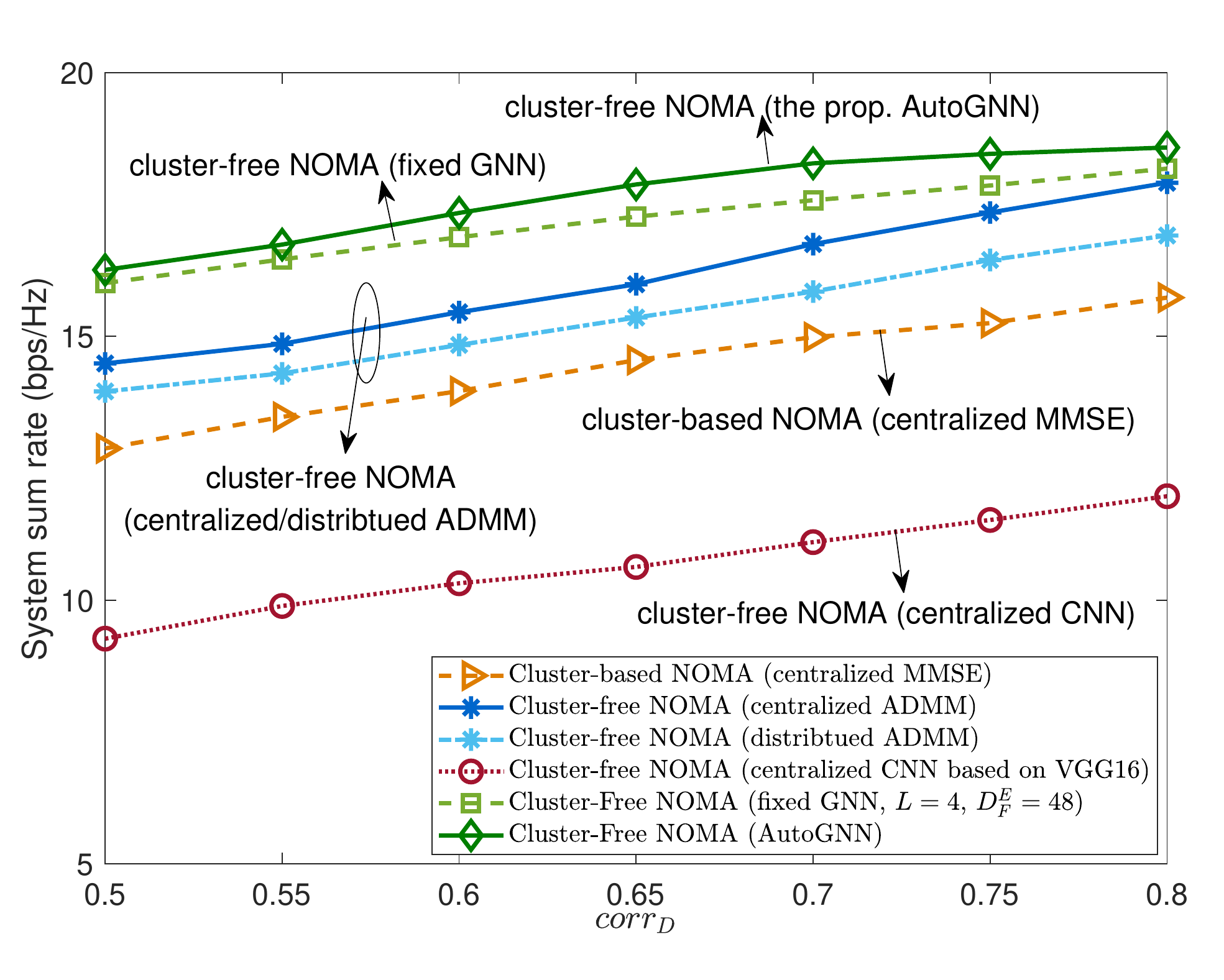}} }\\ \vspace{-0.2em}
    \subfloat[Information overheads. $M=3$, $N_{\mathrm{T}}=4$, $K=6$.]{\centering \scalebox{0.4}{\includegraphics{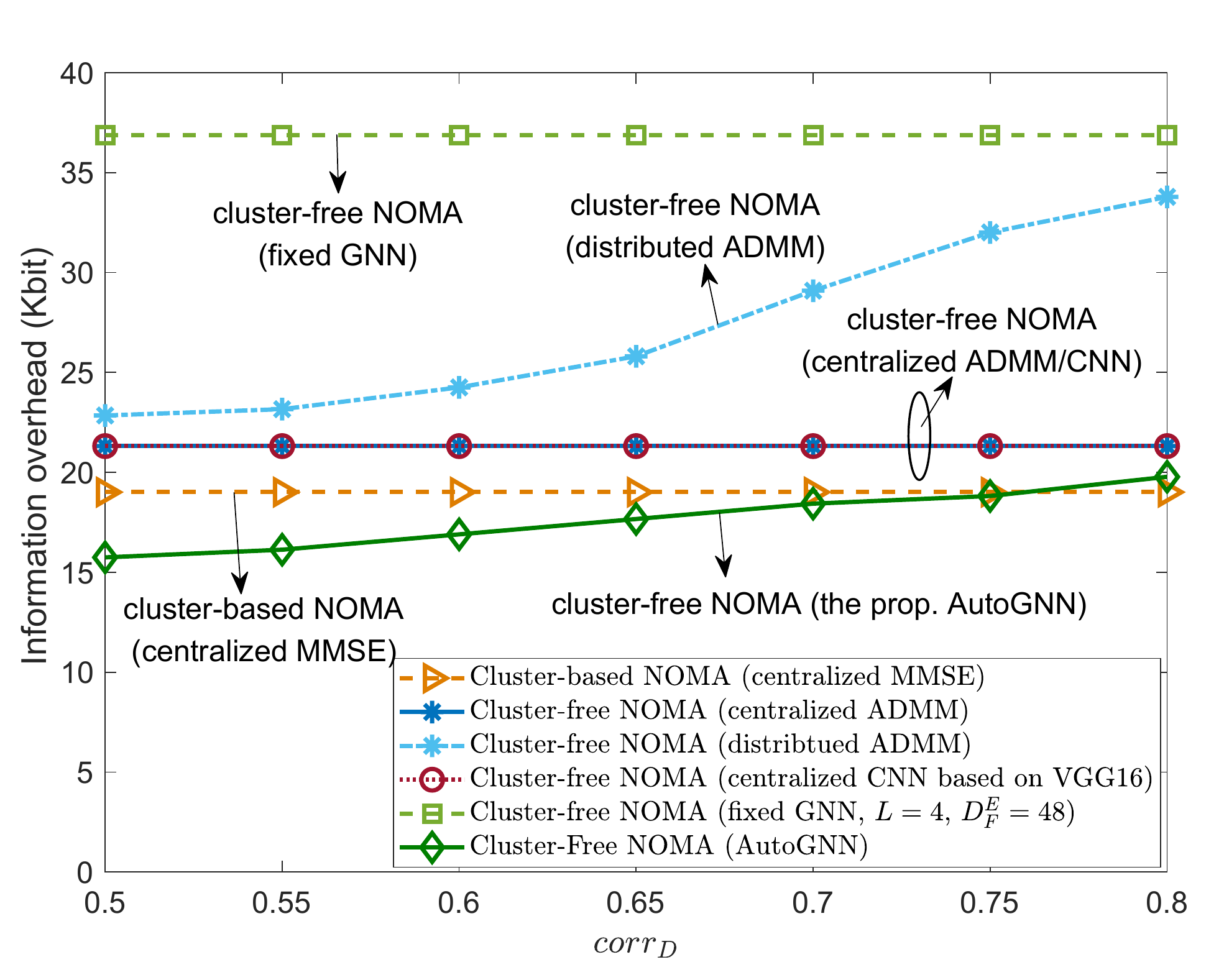}}}\\ \vspace{-0.2em}
    \subfloat[SIC decoding complexity. $M=3$, $N_{\mathrm{T}}=4$, $K=6$.]{\centering \scalebox{0.4}{\includegraphics{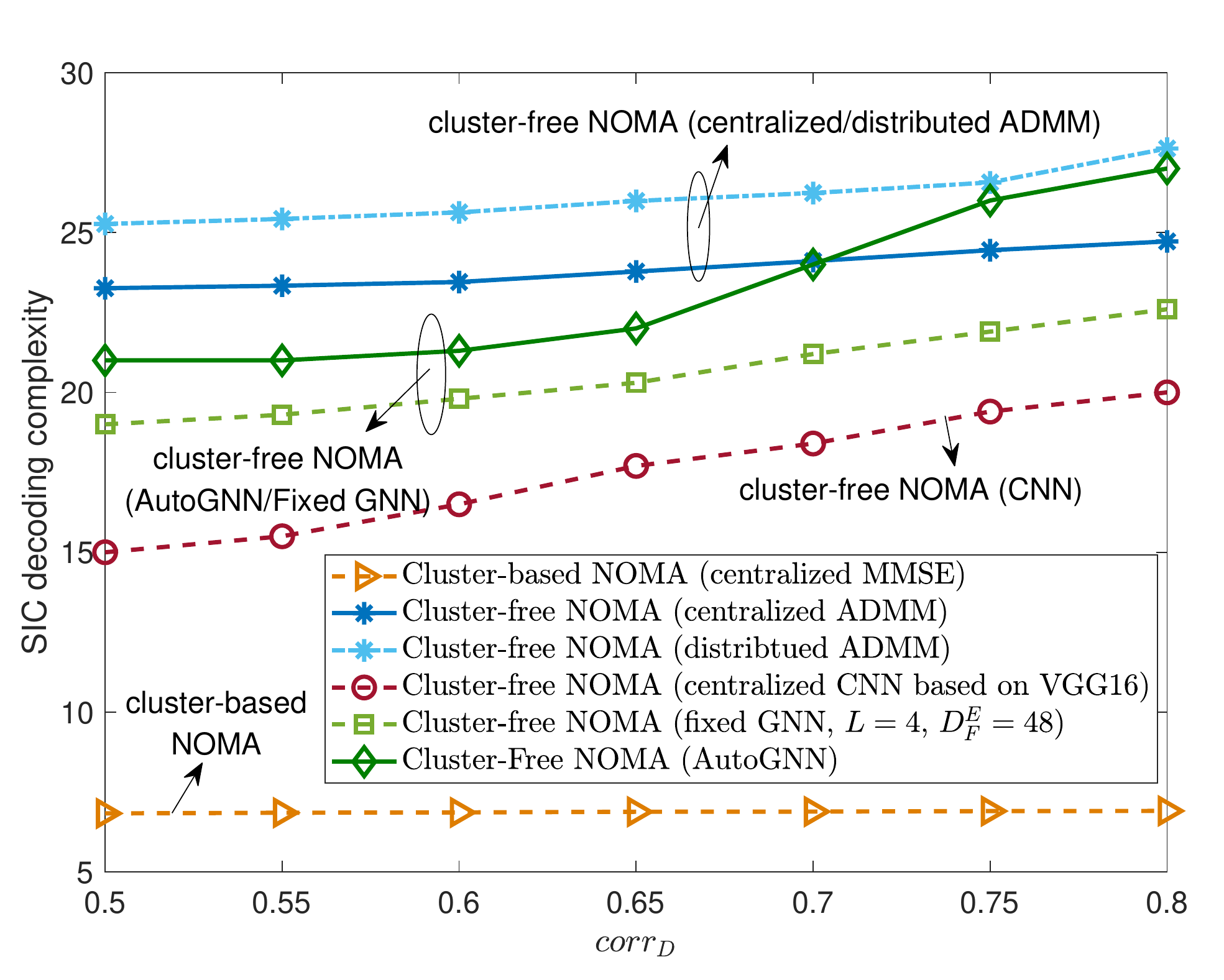}}}
    \caption{System performance comparisons under different channel correlations $corr_{D}$.}\label{fig_corr}
  \vspace{-2em}
\end{figure}

In Fig. \ref{fig_conv}, we compare the convergence behaviours for training the AutoGNN and the benchmarks. 
Detailed numerical results are presented in Table \ref{table:comp}. 
To confirm the effectiveness, we further introduce another automated GNN algorithm into benchmarks, 
namely \textbf{Auto}mated \textbf{G}NN with \textbf{E}xplicit \textbf{L}ink information (AutoGEL) \cite{AutoGEL}, 
which aims to search the optimal functions of embedding, aggregation, combination, activation, and pooling.
Moreover, we also present the performances of two optimization-based algorithms, namely the distributed ADMM and the centralized ADMM algorithms. 
For the centralized/distributed ADMM algorithm, we select the initialized parameters by testing $20$ different initializations.
Fig. \ref{fig_conv} shows both the fixed GNN and the AutoGNN achieve comparable system sum rate with the centralized ADMM algorithm and outperform the distributed ADMM algorithm, 
which validates that the GNN-based algorithms can effectively overcome the parameter initialization dependence issue and achieves efficient coordination.
Moreover, the non-structural and centralized CNN yields the worst performance since it suffers poor scalability in the multi-cell systems. 
From Table \ref{table:comp}, the centralized ADMM achieves low information overheads at the cost of the longest execution time. 
While the distributed ADMM can overcome the slow convergence of conventional optimization-based methods, 
it causes high communication overheads since information should be exchanged during each iteration. 
In contrast, the AutoGNN can obtain comparable performance to AutoGEL while yielding the fastest time response and significantly reducing the information overheads, 
which verifies that it can provide a computation- and communication-efficient paradigm for distributed scheduling.

Fig. \ref{fig_corr} presents the system performance comparisons among different algorithms under various data channel correlations. 
Besides the proposed benchmarks, we also introduce the conventional cluster-based NOMA mechanism to verify the performance gains of the proposed cluster-free NOMA scheme. 
Specifically, the cluster-based NOMA mechanism performs the user clustering based on the channel correlations \cite{MMSE_Dai}, 
and the beamforming vectors of users are optimized in a centralized way based on the MMSE reformulation. 
Here, the results of all algorithms are averaged over the test dataset.

In Fig. \ref{fig_corr}(a), the proposed multi-cell cluster-free NOMA framework outperforms cluster-based NOMA under different data channel correlations, 
and the rate performance gap generally increases with the data channel correlations. 
The centralized ADMM outperforms distributed ADMM in various scenarios, 
which is probably due to the fact that distributed ADMM requires more initialization parameters to be set. 
Since these parameters are difficult to be properly configured in practice, distributed ADMM is more susceptible to the parameter initialization dependence.
The learning-based GNN algorithms outperform conventional distributed ADMM algorithms, 
which demonstrates the efficiency of GNNs to facilitate multi-agent interaction and coordination.
Moreover, owing to the automated architecture optimization capabilities, the proposed AutoGNN yields higher system sum rate than conventional fixed GNN,  
and achieves comparable performance with the well-tuned centralized ADMM algorithm.

Fig. \ref{fig_corr}(b) shows the information overheads required by different algorithms for distributed scheduling.
As shown in Fig. \ref{fig_corr}(b), the distributed ADMM algorithm suffers high information overheads due to the slow convergence.
Moreover, the communication overheads of the proposed AutoGNN increase with data channel correlations, 
which may due to that higher data channel correlations lead to higher SIC decoding complexity and requires more sophisticated distributed control.
Compared with the conventional distributed ADMM and the fixed GNN, 
the proposed AutoGNN can significantly and adaptively reduces the information overheads under different data channel correlations, 
which demonstrates the effectiveness of the auto-learning architecture.

On the other hand, Fig. \ref{fig_corr}(c) compares the SIC decoding complexity, 
i.e., $\frac{1}{N_{\mathrm{sample}}}\sum_{m\in\mathcal{M}}\sum_{i\in\mathcal{K}}\sum_{k\in\mathcal{K},i\ne k}\beta_{ik}^{m}$, between different algorithms, 
with $N_{\mathrm{sample}}$ being the number of samples in the test dataset.
From Fig. \ref{fig_corr}(c), the proposed framework leads to higher SIC decoding complexity than conventional cluster-based NOMA, 
and the SIC decoding complexity adaptively increases with users' data channel correlations. 
This verifies that the proposed framework has a higher flexibility to deal with different scenarios.
Moreover, the SIC decoding complexity of AutoGNN can approach the centralized ADMM algorithm  better than the fixed GNN.

\begin{figure}[!t]
  \vspace{-1em}
    \centering
    \subfloat[System sum rate. $M=5$, $N_{\mathrm{T}}=6$, $K=10$. ]{\centering \scalebox{0.4}{\includegraphics{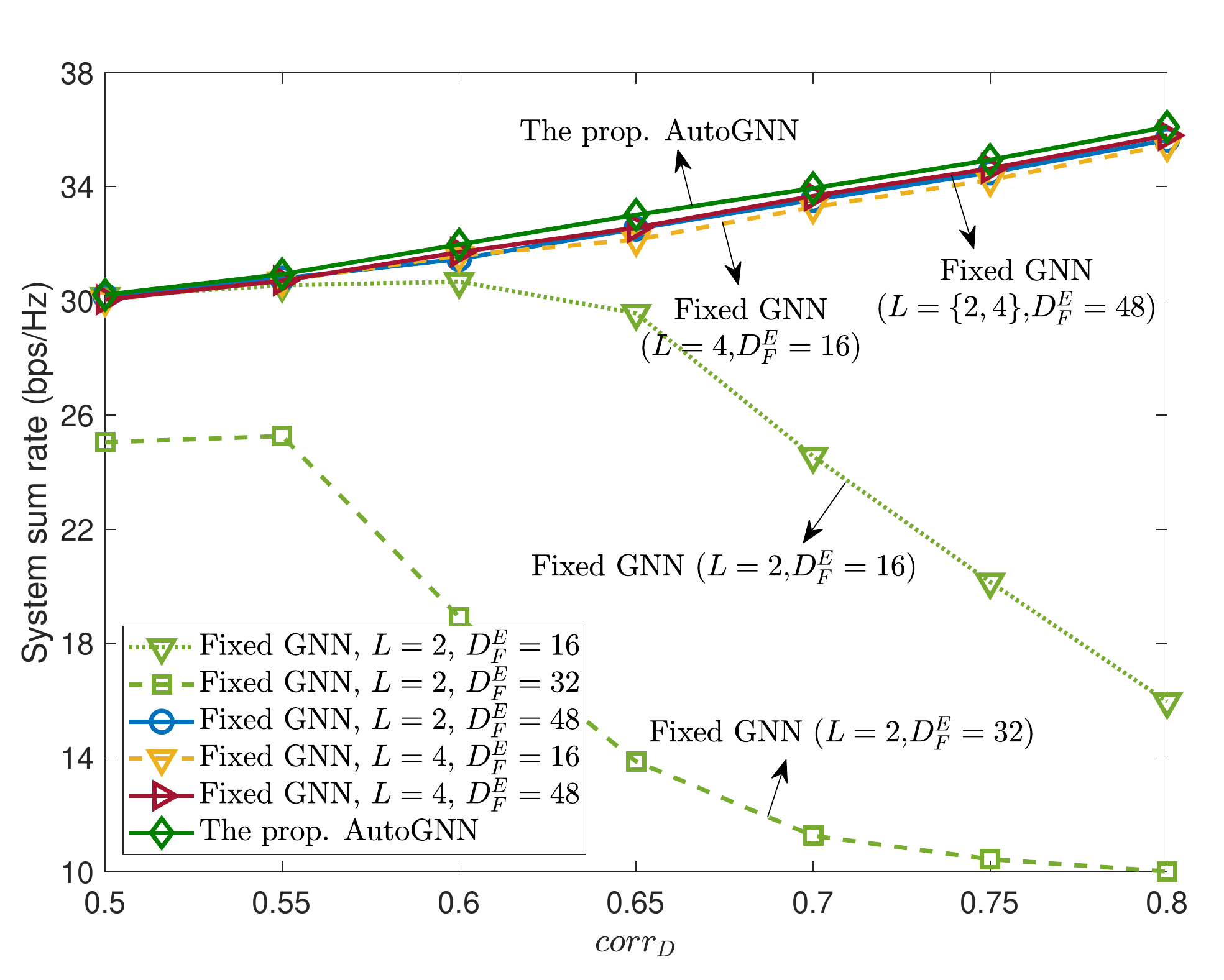}} }\\
    \subfloat[Information overheads. $M=5$, $N_{\mathrm{T}}=6$, $K=10$.]{\centering \scalebox{0.4}{\includegraphics{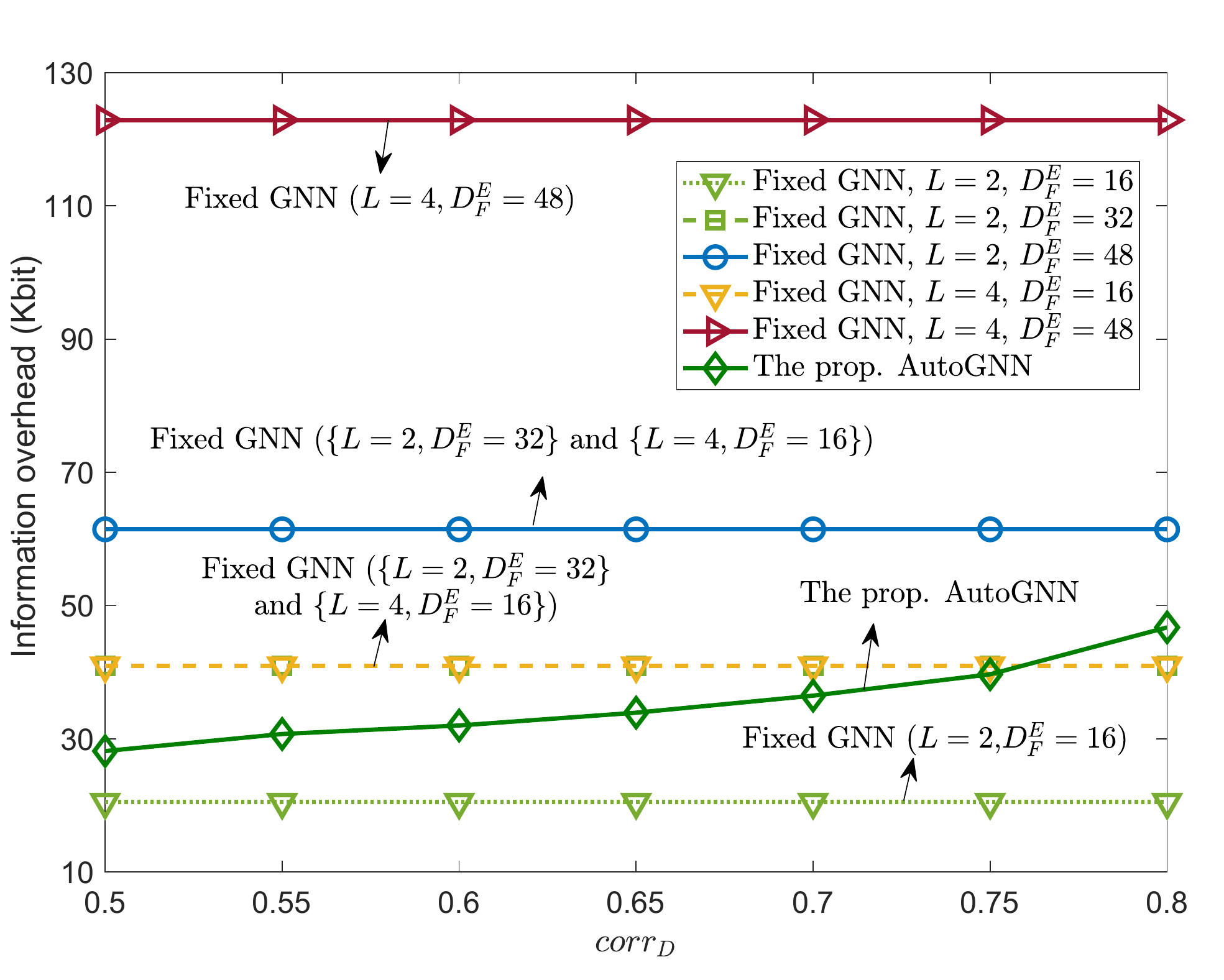}} }
    \caption{Scalability comparisons under different channel correlations $corr_{D}$.}\label{fig_scalibility}
  \vspace{-1em}
\end{figure}

\begin{table*}[!t]
  \begin{minipage}{1\linewidth}
      \centering
      \caption{Generalization from $SNR=20$ dB to $SNR=15$ dB. $corr_{D}=0.7$.} \vspace{-0.2em}
      \label{table:generalization}
      \resizebox{1\textwidth}{!}{
      \begin{tabular}{l|c|c|c|c}
          \hline
             \multirow{2}{*}{Method} & \multicolumn{2}{c|}{$M=3$, $N_{\mathrm{T}}=4$, $K=6$}& \multicolumn{2}{c}{$M=5$, $N_{\mathrm{T}}=6$, $K=10$} \\  \cline{2-5}
              & Sum rate (generalization) & Sum rate (training) & Sum rate (generalization) & Sum rate (training) \\ \hline
             Fixed GNN ($L=4$, $D_{F}^{E} =48$)	& 16.44 bps/Hz	& 16.63	bps/Hz& 32.04 bps/Hz& 32.06 bps/Hz\\ \hline
             AutoGNN &	17.27 bps/Hz&	17.64 bps/Hz& 32.38	bps/Hz& 33.78 bps/Hz\\ \hline
      \end{tabular}
      }
\end{minipage}
\hfill
\vspace{-1.5em}
\end{table*}
  
To demonstrate the scalability performance of the proposed algorithm, in Fig. \ref{fig_scalibility} we further consider a larger network setting with $M=5$, $N_{\mathrm{T}}=6$, and $K=10$.
To verify the effectiveness of AutoGNN, we also present the performance of fixed GNNs with different configurations of network depths and message embedding sizes. 
From Fig. \ref{fig_scalibility}(a), the performance achieved by fixed GNNs will degrade dramatically when reducing the message embedding size and/or the network depth. 
However, when $L=2$, the performance of fixed GNN with $D_{F}^{E}=16$ can outperform the one with $D_{F}^{E}=32$, 
which signifies that the message embedding size should be appropriately  configured. 
Moreover, the fixed GNNs with configurations $\left\{L=4, D_{F}^{E}=16 \right\}$, $\left\{L=2, D_{F}^{E}=48 \right\}$, 
and $\left\{L=4, D_{F}^{E}=48 \right\}$ can achieve similar representation capabilities. 
In comparison, 
the proposed AutoGNN can automatically optimize and adaptively configure the GNN structure to obtain a communication-efficient architecture, 
which outperforms the fixed GNNs built on pre-defined structures. 
As shown in Fig. \ref{fig_scalibility}(b), while the fixed GNN with $\left\{L=2, D_{F}^{E}=16 \right\}$ yielding the lowest information overheads, 
the achieved system sum rate dramatically decreases with channel correlations. 
Moreover, the fixed GNNs with $\left\{L=2, D_{F}^{E}=48 \right\}$, $\left\{L=4, D_{F}^{E}=16 \right\}$, 
$\left\{L=4, D_{F}^{E}=48 \right\}$ can ensure the system performance at the cost of high communication overheads. 
In contrast, the proposed AutoGNN can always achieve the highest performance while keeping low communication burdens intelligently, despite the  increasing network scale. 
This validates the superiority of the proposed self-optimized AutoGNN in terms of the representation ability. 

Table \ref{table:generalization} verifies the generalization ability of GNNs trained from the setting of $SNR=20$ dB to an unseen scenario of $SNR=15$ dB. 
When applying the models trained under $SNR=20$ dB to $SNR=15$ dB, 
both the fixed GNN and the AutoGNN can achieve similar performance compared to the ones that are directly trained under $SNR=15$ dB. 
Moreover, the generalization ability of the learned models may not be compromised when the wireless network scale increases.

\section{Conclusion}
A novel multi-cell cluster-free NOMA framework has been proposed in this paper, where the coordinated beamforming 
and cluster-free SIC were jointly designed to suppress both intra-cell and inter-cell interference.
The objective function is formulated to maximize the system sum rate while ensuring SIC decoding constraints and users' minimum rate requirements.
To deal with this highly coupling and complex MINLP problem, a novel communication-efficient distributed AutoGNN architecture was proposed, 
which can automatically tailor the GNN architecture to reduce computation and information exchange burdens.
To jointly train the GNN model weights and architecture parameters for distributed beamforming and SIC optimization, 
a bi-level AutoGNN learning algorithm was developed, which was theoretically proven to converge to a stationary point.
Our numerical results demonstrated that the cluster-free NOMA outperforms conventional cluster-based NOMA under multi-cell scenarios.
Moreover, compared with the conventional fixed GNN and distributed ADMM algorithms, 
the proposed AutoGNN can significantly reduce the computation and communication overheads while optimizing the system performance.  
For future directions, some open challenges remain to be addressed. 
Specifically, how to effectively accelerate and fully decentralize the bi-level training process of AutoGNN remains a crucial issue. 
Furthermore, to reduce overheads of both CSI estimations and scheduling algorithms, end-to-end learning-based schemes are urgently required for practical designs of multi-cell NOMA networks. 

\ifCLASSOPTIONcaptionsoff
  \newpage
\fi

\appendix
\subsection{Proof of \textbf{Lemma} \ref{theorem_Neumman_hypergradient}} \label{proof_theorem_Neumman_hypergradient}
Let $\upsilon\left(\mathbf{G}_{*}\right)$ collect all the eigenvalues of the Hessian matrix $\mathbf{G}_{*}$.
Using condition (iii) in Assumption \ref{Assumption_L1}, the eigenvalues $\upsilon\left(\mathbf{G}_{*}\right)$ are lower bounded by $0<\mu\leq\upsilon\left(\mathbf{G}_{*}\right)$.
Using condition (ii) in Assumption \ref{Assumption_L1}, the eigenvalues $\upsilon\left(\mathbf{G}_{*}\right)$ are upper bounded by $\widetilde{L}^{\theta}$, namely,
$\mathbf{G}_{*}\preceq \widetilde{L}^{\theta}\mathbf{I} \Rightarrow 0 < \upsilon\left(\mathbf{G}_{*}\right) \le \widetilde{L}^{\theta}$.
Since $\upsilon\left(\kappa\mathbf{G}_{*}\right)=\kappa\left[\upsilon\left(\mathbf{G}_{*}\right)\right]$, given a learning rate $\kappa<\frac{2}{\widetilde{L}^{\theta}}$, 
we have $0<\kappa\left[\upsilon\left(\mathbf{G}_{*}\right)\right]<\kappa \widetilde{L}^{\theta}<2$, which yields
\vspace{-0.3em}
\begin{equation}\label{eig_kG-I}\small
-1<\upsilon\left(\kappa\mathbf{G}_{*}-\mathbf{I}\right)=\nu\left(\kappa\mathbf{G}_{*}\right)-1<1.
\vspace{-0.3em}
\end{equation}
Since the spectral norm $\Vert\kappa\mathbf{G}_{*}-\mathbf{I}\Vert = \Vert\mathbf{I}-\kappa\mathbf{G}_{*}\Vert = \max\left\{\left|\upsilon\left(\kappa\mathbf{G}_{*}-\mathbf{I}\right)\right|\right\}$, from \eqref{eig_kG-I} we can obtain
$\Vert\mathbf{I}-\kappa\mathbf{G}_{*}\Vert <1$.
Therefore, according to \textbf{Lemma} \ref{lemma_Neumann}, $\mathbf{G}_{*}^{-1}$ can be approximated by the Neumann series expansion \eqref{Neumann} as
\vspace{-0.3em}
\begin{equation}\label{NeumannDerivative}\small
\mathbf{G}_{*}^{-1}
=\kappa\left(\kappa\mathbf{G}_{*}\right)^{-1}
=\kappa\sum_{n=0}^{\infty}\left(\mathbf{I}-\kappa\mathbf{G}_{*}\right)^{n}.
\vspace{-0.3em}
\end{equation}
Substituting \eqref{NeumannDerivative} into \eqref{ImplicitHypergradient}, the implicit hypergradient defined in \eqref{ImplicitHypergradient} can be rewritten as \eqref{HyperGradient_Neumann}, which ends the proof.

\vspace{-0.3em}
\subsection{Proof of \textbf{Theorem} \ref{theorem_convergence}} \label{proof_theorem_convergence}
We first demonstrate that the bi-level learning algorithm yields a non-decreasing sequence of training losses, which can reach convergence during the training process. 
Thereafter, its convergence to the stationary point is proven.

\textit{\textbf{(i) Non-decreasing sequence of training losses:}}
Let $\nabla_{\bm{\alpha}}\mathcal{L}_{\mathrm{v}}^{u} \! \triangleq \! \nabla_{\bm{\alpha}}\mathcal{L}_{\mathrm{v}}\left(\bm{\Theta}^{u},\bm{\alpha}^{u}\right)$ 
and $\nabla_{\bm{\alpha}}\overline{\mathcal{L}}_{\mathrm{v}}^{iu} \! \triangleq \! \overline{\mathcal{L}}_{\mathrm{v}}^{i}\left(\bm{\Theta}^{u},\bm{\alpha}^{u}\right)$
denote the exact hypergradient and the non-approximated stochastic hypergradient with $N_{\mathrm{G}}\to\infty$ at each outer-loop iteration $u$, respectively.
Define $\varepsilon^{u}\triangleq \nabla_{\bm{\alpha}}\mathcal{L}_{\mathrm{v}}^{u}
-\nabla_{\bm{\alpha}}\overline{\mathcal{L}}_{\mathrm{v}}^{iu}$ as the noise between the exact hypergradient and the non-approximate stochastic hypergradient.
According to \cite{iDARTS}, the stochastic gradient $\nabla_{\bm{\alpha}}\overline{\mathcal{L}}_{\mathrm{v}}^{iu}$ provides an unbiased estimate of $\nabla_{\bm{\alpha}}\mathcal{L}_{\mathrm{v}}^{u}$,
i.e., 
\vspace{-0.3em}
\begin{equation}\label{error_unbiased}\small
\mathbb{E}\left[\varepsilon^{u}\right]\triangleq\mathbb{E}\left[
  \nabla_{\bm{\alpha}}\mathcal{L}_{\mathrm{v}}^{u} - 
  \nabla_{\bm{\alpha}}\overline{\mathcal{L}}_{\mathrm{v}}^{iu}
\right]=0,
\vspace{-0.3em}
\end{equation}
where the expectation is taken over all mini-batches.
Denote $\nabla_{\bm{\alpha}}\widetilde{\mathcal{L}}_{\mathrm{v}}^{u} = \nabla_{\bm{\alpha}}\widetilde{\mathcal{L}}_{\mathrm{v}}\left(\bm{\Theta}^{u},\bm{\alpha}^{u}\right)$.
From the definitions we can obtain
\vspace{-0.3em}
\begin{equation}\label{relation_ship_1}\small
\mathbb{E}\!\left[\nabla_{\bm{\alpha}}\mathcal{L}_{\mathrm{v}}^{u}\right]
\overset{\eqref{appr_error}}{=}
\mathbb{E}\!\left[\nabla_{\bm{\alpha}}\widetilde{\mathcal{L}}_{\mathrm{v}}^{u}\right]
\!+\!\mathbb{E}\!\left[\delta^{u}\right]\!,
\vspace{-0.3em}
\end{equation}
\begin{equation}\label{relation_ship_2}\small
  \mathbb{E}\left[\nabla_{\bm{\alpha}}\widehat{\mathcal{L}}_{\mathrm{v}}^{u}\big|\bm{\alpha}^{u}\right]=\mathbb{E}\left[\nabla_{\bm{\alpha}}\widetilde{\mathcal{L}}_{\mathrm{v}}^{u}
  -\varepsilon_{n}\big|\bm{\alpha}_{n}\right]
  \!\overset{\small\eqref{error_unbiased}}{=}\!
  \mathbb{E}\left[\nabla_{\bm{\alpha}}\widetilde{\mathcal{L}}_{\mathrm{v}}^{u}\right].
  \vspace{-0.3em}
\end{equation}
\begin{equation}\label{relation_ship_3}\small
  \mathbb{E}\!\!\left[\!\left\Vert \nabla_{\bm{\alpha}}\widehat{\mathcal{L}}_{\mathrm{v}}^{u}\!\right\Vert ^{2}\!\right]
\!\!=\!\!\mathbb{E}\!\!\left[\!\left\Vert \nabla_{\bm{\alpha}}\widetilde{\mathcal{L}}_{\mathrm{v}}^{u}
\!-\!\varepsilon^{u}\!\right\Vert ^{2}\!\right]
\!\overset{\small\eqref{error_unbiased}}{=}
\!\mathbb{E}\!\left[\left\Vert\! \nabla_{\bm{\alpha}}\widetilde{\mathcal{L}}_{\mathrm{v}}^{u}\!\right\Vert ^{2}\!\right]\!
\!+\!\mathbb{E}\!\!\left[\!\left\Vert \varepsilon^{u}\!\right\Vert ^{2}\!\right]\!.
\vspace{-0.3em}
\end{equation}
Furthermore, since function $\nabla_{\bm{\alpha}}\mathcal{L}_{\mathrm{v}}^{u}$ is Lipschitz continuous with constant $\widetilde{L}_{\mathrm{v}}^{\alpha}$, according to the Lipschitz condition we have
\vspace{-0.3em}
\begin{equation*}\small
\begin{split}
& 
\mathbb{E}\!\left[\mathcal{L}_{\mathrm{v}}^{u+1}\right]
\!\le\!
\mathbb{E}\!\left[\mathcal{L}_{\mathrm{v}}^{u}\right]
\!+\!\mathbb{E}\!\left[\!\left\langle \nabla_{\bm{\alpha}}\mathcal{L}_{\mathrm{v}}^{u},\bm{\alpha}^{u+1}\!\!-\!\bm{\alpha}^{u}\right\rangle \! \right]
\!+\!\frac{\widetilde{L}_{\mathrm{v}}^{\alpha}}{2}\mathbb{E}\!\left[\!\left\Vert \bm{\alpha}^{u+1}\!\!-\!\bm{\alpha}^{u}\right\Vert ^{2}\!\right]
\\&
=\mathbb{E}\!\left[\mathcal{L}_{\mathrm{v}}^{u}\right]
\!+\!\left\langle\! \mathbb{E}\!\left[\nabla_{\bm{\alpha}}\mathcal{L}_{\mathrm{v}}^{u}\right],
\!-\!\kappa^{u}\mathbb{E}\!\left[\!\nabla_{\bm{\alpha}}\widehat{\mathcal{L}}_{\mathrm{v}}^{u}\!\right]\!\right\rangle 
\!+\!\frac{\widetilde{L}_{\mathrm{v}}^{\alpha}}{2}\!\left(\kappa^{u}\right)^{2}\!\mathbb{E}\!\left[\left\Vert\! \nabla_{\bm{\alpha}}\widehat{\mathcal{L}}_{\mathrm{v}}^{u}\!\right\Vert\!^{2}\!\right]\!.
\end{split}
\vspace{-0.3em}
\end{equation*}
Substituting \eqref{relation_ship_1}-\eqref{relation_ship_3} into the above equation we have
\vspace{-0.3em}
\begin{equation}\label{expectation_L_2}\small
\begin{split}
\mathbb{E}\!\left[\mathcal{L}_{\mathrm{v}}^{u+1}\right]
&\!\!\le\!
\mathbb{E}\!\left[\mathcal{L}_{\mathrm{v}}^{u}\right]\!-\!\!\kappa^{u}\mathbb{E}\!\left[\left\Vert\nabla_{\bm{\alpha}}\widetilde{\mathcal{L}}_{\mathrm{v}}^{u}\right\Vert^{2}\!\right]\!
\!-\!\kappa^{u}\mathbb{E}\!\!\left[\!\left\langle \delta^{u}\!,\!\nabla_{\bm{\alpha}}\widetilde{\mathcal{L}}_{\mathrm{v}}^{u}\right\rangle \right]
\\&
\!+\!\!\frac{\widetilde{L}_{\mathrm{v}}^{\alpha}}{2}\!\left(\kappa^{u}\right)^{2}\mathbb{E}\!\!\left[\left\Vert \nabla_{\bm{\alpha}}\widetilde{\mathcal{L}}_{\mathrm{v}}^{u}\right\Vert ^{2}\right]\!
\!+\!\frac{\widetilde{L}_{\mathrm{v}}^{\alpha}}{2}\!\left(\kappa^{u}\right)^{2}
\!\mathbb{E}\!\left[\left\Vert \varepsilon^{u}\right\Vert ^{2}\right]\!.
\end{split}
\end{equation}
\vspace{-1em}

Using the inequality $\left\Vert \delta^{u}\right\Vert \!\le \! C_{\mathrm{v}}^{\theta}C^{\theta\alpha}\frac{1}{\mu}\left(1\!-\!\kappa\mu\right)^{N_{\mathrm{G}}+1}$ from \textbf{Lemma} \ref{lemma_error},
we can obtain 
$\left\langle \delta^{u},\nabla_{\bm{\alpha}}\widetilde{\mathcal{L}}_{\mathrm{v}}^{u}\right\rangle \! \ge \! -\Omega\left\Vert \nabla_{\bm{\alpha}}\widetilde{\mathcal{L}}_{\mathrm{v}}^{u}\right\Vert ^{2}$
 with $\Omega \!\triangleq\! \frac{C_{\mathrm{v}}^{\theta}C^{\theta\alpha}\left(1-\kappa\mu\right)^{N_{\mathrm{G}}+1}}{\mu\left\Vert \nabla_{\bm{\alpha}}\widetilde{\mathcal{L}}_{\mathrm{v}}^{u}\right\Vert }$.
Since we can generally assume 
$\mathbb{E}\left[\left\Vert \varepsilon^{u}\right\Vert ^{2}\right]\leq \eta\left\Vert \nabla_{\bm{\alpha}}\widetilde{\mathcal{L}}_{\mathrm{v}}^{u}\right\Vert ^{2}$  
in the stochastic bi-level gradient learning algorithm  \cite{iDARTS,Reversemode_2018}, 
\eqref{expectation_L_2} can be further recast as
\vspace{-0.3em}
\begin{equation}\small\label{expectation_L_3}
\mathbb{E}\!\!\left[\mathcal{L}_{\mathrm{v}}^{u+1}\right] 
\!\le\!\mathbb{E}\!\left[\mathcal{L}_{\mathrm{v}}^{u}\right]
\!-\!\kappa^{u}\!\left[1\!-\!\Omega\!-\!\frac{\widetilde{L}_{\mathrm{v}}^{\alpha}}{2}\!\kappa^{u}\!\left(1\!+\!\eta\right)\!\right]
\!\mathbb{E}\!\left[\left\Vert \nabla_{\bm{\alpha}}\widetilde{\mathcal{L}}_{\mathrm{v}}^{u}\right\Vert ^{2}\!\right].
\vspace{-0.3em}
\end{equation}
Given that the learning rate is small enough, i.e., $0<\kappa^{u}<\min\left\{\frac{1}{\mu},\frac{2\left(1-\Omega\right)}{\widetilde{L}_{\mathrm{v}}^{\alpha}\left(1+\eta\right)}\right\}$, we can ensure $0<1-\kappa^{u}\mu<1$. 
Hence, by choosing an appropriate $N_{\mathrm{G}}$, we can guarantee $\Omega\ll 1$ and $1-\Omega-\frac{\widetilde{L}_{\mathrm{v}}^{\alpha}}{2}\kappa^{u}\left(1+\eta\right)>0$, which implies that
\vspace{-0.3em}
\begin{equation}\small
\mathbb{E}\left[\mathcal{L}_{\mathrm{v}}^{u+1}\right]\le\mathbb{E}\left[\mathcal{L}_{\mathrm{v}}^{u}\right].
\vspace{-0.3em}
\end{equation}
Since $\mathcal{L}_{\mathrm{v}}$ is bounded due to the limited transmit power and the mutual interference,
it can be decreased by the outer loop updates until reach convergence.

\textbf{\textit{(ii) Convergence to the stationary point:}}
Rearrange \eqref{expectation_L_3} as
\vspace{-0.3em}
\begin{equation}\small\label{convergence_decreasing}
\begin{split}
\mathbb{E}\!\left[\mathcal{L}_{\mathrm{v}}^{u}\right]\!-\!\!\mathbb{E}\left[\mathcal{L}_{\mathrm{v}}^{u+1}\right]
\!\ge\!\kappa^{u}\!\left[1\!-\!\Omega\!-\!\frac{\widetilde{L}_{\mathrm{v}}^{\alpha}}{2}\kappa^{u}\!\left(1\!+\!\eta\right)\!\right]
\!\mathbb{E}\!\left[\!\left\Vert \nabla_{\bm{\alpha}}\widetilde{\mathcal{L}}_{\mathrm{v}}^{u}\right\Vert ^{2}\!\right]\!.
\end{split}
\vspace{-0.3em}
\end{equation}
Summing \eqref{convergence_decreasing} over $u=0,1,...,T_{\mathrm{out}}$, we have
\vspace{-0.3em}
\begin{equation*}\small
\mathbb{E}\!\left[\mathcal{L}_{\mathrm{v}}^{(0)}\right]
\!-\!\mathbb{E}\!\left[\mathcal{L}_{\mathrm{v}}^{T_{\mathrm{out}}}\right]
\!\!\ge\!\!
\sum_{u=0}^{T_{\mathrm{out}}}\!\kappa^{u}\!\!\left[\!1\!-\!\Omega\!-\!\frac{\widetilde{L}_{\mathrm{v}}^{\alpha}}{2}\kappa^{u}\!\!\left(1\!+\!\eta\right)\!\right]
\!\mathbb{E}\!\!\left[\left\Vert \nabla_{\bm{\alpha}}\widetilde{\mathcal{L}}_{\mathrm{v}}^{u}\right\Vert ^{2}\!\right]\!.
\vspace{-0.3em}
\end{equation*}
Since $\mathcal{L}_{\mathrm{v}}$ is lower bounded, we can obtain
\vspace{-0.3em}
\begin{equation*}
\lim\limits_{T_{\mathrm{out}}\to\infty}\!\sum\limits_{u=0}^{T_{\mathrm{out}}}\kappa^{u}\!\!\left[1\!-\!\Omega\!-\!\frac{\widetilde{L}_{\mathrm{v}}^{\alpha}}{2}\kappa^{u}\!\left(1\!+\!\eta\right)\right]
\!\mathbb{E}\!\left[\left\Vert \nabla_{\bm{\alpha}}\widetilde{\mathcal{L}}_{\mathrm{v}}^{u}\right\Vert ^{2}\right]\!<\!\infty.
\vspace{-0.3em}
\end{equation*}
When learning rate $\kappa^{u}$ in each outer-loop iteration $u$ satisfies $\sum_{u=1}^{\infty} \kappa^{u} = \infty$ and $\sum_{u=1}^{\infty} \left(\kappa^{u}\right)^{2} < \infty$ \cite{StepSize_2008},
we have $\lim\limits_{T_{\mathrm{out}}\to\infty}\sum\limits_{n=0}^{T_{\mathrm{out}}}\kappa^{u}\left[1\!-\!\Omega\!-\!\frac{\widetilde{L}_{\mathrm{v}}^{\alpha}}{2}\kappa^{u}\left(1+\eta\right)\right]=\infty$, 
which signifies $\lim\limits_{u\to\infty}\mathbb{E}\left[\left\Vert \nabla_{\bm{\alpha}}\widetilde{\mathcal{L}}_{\mathrm{v}}^{u}\right\Vert ^{2}\right]=0$.
This completes the proof.

\ifCLASSOPTIONcaptionsoff
  \newpage
\fi

\end{document}